\documentclass[a4paper,12pt]{elsarticle}

\usepackage{url}
\usepackage{graphicx}
\usepackage{amssymb, amsmath, amsfonts, amsthm}
\usepackage{multirow}
\usepackage{bm}
\usepackage{nomencl}
\usepackage{etoolbox}
\usepackage{physics}
\usepackage{xstring}
\usepackage{setspace}
\usepackage{acronym}
\usepackage{anysize}
\usepackage[margin=3cm]{geometry}
\usepackage[dvipsnames]{xcolor}
\usepackage{xcolor}
\usepackage{colortbl}
\usepackage{caption}
\usepackage{hyperref}
\hypersetup{
    hidelinks,
    colorlinks=false,
    urlcolor=black,
    linkcolor=black,
    citecolor=black
}

\usepackage{titlesec}
\titlespacing*{\section}{0pt}{8pt plus 2pt minus 2pt}{4pt plus 1pt minus 1pt}
\titlespacing*{\subsection}{0pt}{6pt plus 2pt minus 2pt}{2pt plus 1pt minus 1pt}

\makeatletter
\let\MYcaption\@makecaption
\makeatother

\journal{Electric Power Systems Research}

\let\oldtabular\tabular
\renewcommand{\tabular}{\footnotesize\oldtabular}

\usepackage[font=footnotesize]{subcaption}
\makeatletter
\let\@makecaption\MYcaption
\makeatother

\ExplSyntaxOn

\ExplSyntaxOff

\renewcommand\nomgroup[1]{%
  \item[\bfseries
  \ifstrequal{#1}{A}{Operators}{%
  \ifstrequal{#1}{B}{Subscripts}{%
  \ifstrequal{#1}{C}{General Variables}{%
  \ifstrequal{#1}{D}{Electrical Variables}{%
  }}}}%
]}

\acrodef{gfm}[GFM]{Grid-Forming}
\acrodef{gfl}[GFL]{Grid-Following}
\acrodef{vs}[VS]{Voltage Source}
\acrodef{cs}[CS]{Current Source}
\acrodef{gnc}[GNC]{Generalized Nyquist Criterion} 
\acrodef{nfp}[NFP]{Network Frequency Perturbation} 
\acrodef{pll}[PLL]{Phase-Locked Loop}
\acrodef{srf}[SRF]{Synchronous Reference Frame}
\acrodef{pcc}[PCC]{Point of Common Coupling}

\marginsize{2.54cm}{2.54cm}{2.54cm}{2.54cm}

\makenomenclature

\begin{document}

\setlength{\abovedisplayskip}{5pt}  
\setlength{\belowdisplayskip}{5pt}  
\setlength{\abovedisplayshortskip}{5pt}
\setlength{\belowdisplayshortskip}{5pt}

\setlength{\bibsep}{0pt plus 0.3ex}

\begin{frontmatter}

\allowdisplaybreaks

\title{Equivalent modelling for the fundamental frequency dynamic variation: State-space, impedance, and power-frequency representations\tnoteref{cr}}

\tnotetext[cr]{\textcopyright{} 2026. This manuscript version is made available under the CC-BY-NC-ND 4.0 license \url{https://creativecommons.org/licenses/by-nc-nd/4.0/}. The Version of Record is available at \url{https://doi.org/10.1016/j.epsr.2026.113098}}
\author[inst2]{Pablo Rodríguez-Ortega}
\ead{pablo.rodriguez@imdea.org}

\author[inst3]{Dionysios Moutevelis}
\ead{dionysios.moutevelis@upc.edu}

\author[inst2,inst4]{Javier Roldán-Pérez}
\ead{javier.roldan@imdea.org}

\author[inst2]{Milan Prodanovic}
\ead{milan.prodanovic@imdea.org}

\address[inst2]{Electrical Systems Unit, IMDEA Energy, M\'{o}stoles, Spain}

\address[inst3]{CITCEA-UPC, Barcelona, Spain}

\address[inst4]{Escuela Técnica Superior de Ingenieros Industriales, Universidad Politécnica de Madrid, Spain\vspace{-0.7cm}}

\setstretch{1.4}
\begin{abstract}
Stability of power electronic converters connected to power grids is commonly assessed by using the impedance criterion while the stability of power grids is typically analysed by using the network state-space representation.
It is known that the impedance criterion may lead to erroneous results if the grid frequency dynamics are not considered while eigenvalue analysis is considered as a reliable method for system stability assessment.
The equivalence between these two methods has been recently explored, {without considering the effect of network frequency variations.}
Additionally, the link of the impedance criterion with the power-frequency dynamics of power systems also remains largely unexplored.
In this paper, the equivalency between the  impedance method considering the grid frequency dynamics and the conventional eigenvalue analysis is demonstrated.
In addition, the dynamic interaction between the apparent power flow and the network fundamental frequency is formulated and its link with the impedance representation is shown.
It is demonstrated that, by using the impedance representation with the network frequency as an additional input port, the network frequency perturbation plot (NFP) can be intuitively expressed by using quantities consistent with the impedance analysis framework.
The main findings are verified using detailed numerical simulations of two representative systems. 
\end{abstract}

\begin{keyword}
Impedance criterion, state-space representation, eigenvalue analysis, power-frequency dynamics, grid frequency dynamics, grid-forming, grid-following, network frequency perturbation plot.

\vspace{+0.2cm}

This work is supported by the project REDESFUERTES (PID2022-142416OB-I00), funded by MICIU/AEI/10.13039/501100011033 and by the European Union, NextGenerationEU/PRTR. 
The authors also acknowledge the support of project FLEXENER (MIG-20201002), funded by CDTI.
\end{keyword}
\end{frontmatter}

\acresetall 

\setstretch{1.0}
\printnomenclature
\nomenclature[A, 01]{$\Delta$}{Small perturbation}
\nomenclature[A, 02]{$\dot{}$}{Derivative}
\nomenclature[A, 03]{$^{\intercal}$}{Matrix transpose}
\nomenclature[A, 04]{$^{-1}$}{Inverse matrix}

\nomenclature[B, 01]{$V$, $C$}{Voltage Source (VS), Current Source (CS)}
\nomenclature[B, 02]{$SYS$}{System}
\nomenclature[B, 03]{$dq$}{Park components}
\nomenclature[B, 04]{$g$}{Grid}
\nomenclature[B, 05]{$o$}{Steady state value}

\nomenclature[C, 01]{$\boldsymbol{A}$, $\boldsymbol{B}$, $\boldsymbol{C}$, $\boldsymbol{D}$}{State, input, output and feed-forward matrices}
\nomenclature[C, 02]{$\boldsymbol{x}$, $\boldsymbol{u}$, $\boldsymbol{y}$}{State, input and output vectors}
\nomenclature[C, 03]{$\boldsymbol{H}$, $\boldsymbol{L}$}{Characteristic equation and minor loop transfer functions}
\nomenclature[C, 04]{$R(s)$}{Transfer function between fundamental grid frequency and active power injeciton, used for the Network Frequency Perturbation (NFP) plot definition}

\nomenclature[D, 01]{$\boldsymbol{v_{dq}}(t)
$, $\boldsymbol{i_{dq}}(t)
$, $\boldsymbol{V_{dq}}(s)$, $\boldsymbol{I_{dq}}(s)$}{Voltage and current Park vectors in the time and Laplace domains, measured in Volts [V] and Amperes [A]}
\nomenclature[D, 02]{$\boldsymbol{Z}$, $\boldsymbol{Y}$}{Impedance and Admittance transfer function matrices, measured in Ohms [$\Omega$] and Siemens [S]}
\nomenclature[D, 03]{$\omega_V(t)$ and $\Omega_V(s)$}{Fundamental network angular frequency in the time and Laplace domains, measured in [rad/s]}
\nomenclature[D, 04]{$\boldsymbol{\Psi}$, $\boldsymbol{\Gamma}$}{Frequency-admittance and frequency-impedance transfer function vectors, measured in [A/(rad/s)] and [(rad/s)/A]}
\nomenclature[D, 05]{$\omega_V(t)$, $\Omega_V(s)$}{Fundamental network frequency, measured in Hertz [Hz]}
\nomenclature[D, 05]{$P$, $Q$}{Active and reactive power injections, measured in Watts [W] and Voltampere reactive [VAr]}

\setstretch{1.37}

\vspace{-0.2cm}
\section{Introduction}
\label{sec.introduction}
Stability assessment of power systems with high penetration of converter-based resources is an ongoing research problem, with eigenvalue analysis based on state-space models and impedance analysis based on voltage and current transfer functions being two of the most prominent methods~\cite{chen2023small}.
Aside from fast interactions between the different network components on the electromagnetic scale, the integration of a large number of renewable energy sources via power converter interfaces leads to reduced levels of system inertia~\cite{Power_electronic_integration,milano2018foundations}.
Consequently, the grid frequency dynamics have recently gained importance as they have been demonstrated to present
important implications for stability assessment~\cite{Effect_in_FreqDynamics,DFIG_Frequency_Disturbances}.

{The application of eigenvalue analysis from state-space models} requires detailed information about all the system elements, including hardware and control parameters.
This method can be readily applied in conventional power systems since the number of generators is relatively small and their parameters are well-known or simple to estimate based on standard values.
However, most of the newly-commissioned generators are connected to the grid via electronic interfaces, whose modelling is not straightforward {due to the wide variety of existing controllers} and their parameters may remain hidden due to confidentiality restrictions~\cite{mohammed2022comparison,cecati2025interoperability}.

An alternative method for stability assessment in systems dominated by electronic interfaces is the impedance criterion~\cite{sun2011impedance,sun2022frequency,SIMON2022108528}.
With this method, only the frequency response of the grid impedance, as seen from the converter connection point, is required.
The latter can be obtained by using either numerical simulations or measurements from the real power grid, therefore avoiding the need for detailed grid information~\cite{zhou2024rapid}.  
In recent years, this property of the impedance-based stability analysis has motivated researchers to actively contribute to this field, with recent developments focusing on methods for the efficient and accurate impedance estimation using a variety of signals, e.g., single-tone or pseudorandom, and under a variety of conditions, e.g., during network topology changes and unbalances~\cite{zhou2024rapid,NabilFrequencyEstimation,mohammed2022fast,mohammed2019online,haberle2023mimo,chakraborty2022review,gong2020dq,luhtala2018implementation}.
Additionally, recent publications have also reported advanced converter control techniques that consider the impedance estimation before the asset interconnection, as well as for online operation~\cite{mohammed2022online,reissner2024robust,moutevelis2024virtual}. 
One of the drawbacks of the impedance method is that the stability criteria are less obvious and more difficult to evaluate compared to the eigenvalue method, especially for multi-converter systems~\cite{Small-signal_Methods_Review,Review_Small_signal_modelling}.

Several works found in the literature combine elements of these methods or address their equivalence to a different extent~\cite{chen2023small,amin2017small,Nyquist_Eigen_Analisys,rygg2017apparent,zhu2021participation,yang2021siso,zhang2019impedance}.
In~\cite{chen2023small}, a thorough comparison of various stability assessment methods was performed, including state-space and impedance-based approaches, demonstrating their equivalence under certain conditions, e.g., that no unstable zero-pole cancellation takes place.
In~\cite{amin2017small} the impedance and the eigenvalue methods were compared for a system consisting of a two-terminal VSC-HVDC interconnection, demonstrating the equivalence between the two.
In~\cite{Nyquist_Eigen_Analisys}, precision issues of the impedance criterion were revealed by performing stability analysis based on both the eigenvalue and impedance methods for two different cases.
In~\cite{rygg2017apparent}, the authors adapted the classical open-loop impedance analysis approach to a closed-loop format, demonstrating that the poles of the proposed apparent impedance transfer function are equivalent to the observable eigenvalues at the connection point of a converter.
References~\cite{zhu2021participation,yang2021siso} present novel methods to extend the impedance stability analysis by incorporating participation factor analysis, an approach typically associated with state-space methods.
Finally, the effect of the rotation of the impedance and admittance matrices is studied in~\cite{zhang2019impedance}.
This rotation is closely linked with the reference frame selection for the impedance modelling and can be potentially used to incorporate variations in the voltage angle of the network~(and thus also the fundamental network frequency).
However, in~\cite{zhang2019impedance}, the angle deviation between the network and converter reference frames is treated as constant, instead of a dynamic variable, and readily calculated from the static power flow equations.
Remarkably, none of the above references explicitly incorporate the variations of the network fundamental frequency by modelling it as a distinct dynamic variable, which can lead to potential inaccuracies for stability assessment of low-inertia power systems without stiff frequency regulation.

The fundamental frequency dynamic changes are commonly studied in power systems using power-frequency models~\cite{kundur1994power,11305437}, or \ac{nfp} plots~\cite{IET_NFP}.
The latter have recently received renewed attention, since they are now required in some countries for converter grid code compliance~\cite{GC0137}.
An important flaw of the original impedance criterion is that, despite accounting for variations in the frequency operating point, it does not account for the fundamental grid frequency dynamics, leading to inaccurate results regarding low-frequency oscillation phenomena~\cite{Nyquist_Eigen_Analisys,Small-Signal_Stability_Analysis_Based_Measured}.
In~\cite{wei2025unified}, a unified method for analysing harmonic stability, verified by the classical impedance criterion, and frequency/voltage stability was proposed.
However, in this work, the analytic relationship between the proposed method and the standard stability assessment via state-space representation and eigenvalue analysis was not demonstrated. 
Hence, an analytic framework that links all three approaches, namely the eigenvalue analysis from state-space models, the impedance criterion and the power-frequency models that study slower, electromechanical phenomena would be of interest for the combined study of dynamic interactions across multiple timescales.

Several recent publications have shown that it is possible to include the fundamental grid frequency dynamics in the impedance criterion (e.g., by including the \ac{pll} dynamics)~\cite{Impedance-based_Grid_Synchronization,Impedance-Based_Whole_System_Modelling,Impedance-Based_Analysis_DFIG,DFIG_PLL_IJEPES}).
An extended impedance representation that includes the grid frequency dynamics was introduced in~\cite{Boroyevich}.
In this work, the authors presented an identification method to obtain the transfer functions that link the grid frequency dynamics with the electrical variables (current and voltage), termed \emph{frequency-admittance} and \emph{frequency-impedance}, together with a general mathematical formulation that includes the grid frequency dynamics in the impedance stability criterion.  
The results showed that the accuracy for the system stability assessment increased greatly compared to the original impedance criterion.
Recently, it has also been shown that the different reference frames used to model the impedance can lead to similar results if they are chosen appropriately~\cite{XiongfeiReferenceFrame}, being one of the main differences between methods the presence of an additional input put that models the frequency dynamics, independently.
Since then, the extended impedance criterion that includes the grid fundamental frequency dynamics has been used to reveal new instability mechanisms in~\acp{pll}~\cite{wu2024influence}, as well as to improve the control design in microgrid applications~\cite{moutevelis2024virtual,li2022systematic}.

In this paper, the concepts of frequency-admittance and frequency-impedance are used to formulate the analytic relation between three widely used methods for stability assessment, namely the state-space, impedance and power-frequency methods.
To the extent of the Authors' knowledge, the equivalence of these three methods, with the explicit inclusion of the network fundamental frequency variations, has not been explored in the literature.
The contributions of this work can be summarized as follows:
\begin{itemize}
\item The impedance-based stability assessment method including the grid frequency dynamics is analytically compared with the eigenvalue method, for the fully modelled state-space representation of power systems.
\item The conditions for the equivalence between the two methods are derived.
This is one of the main contributions of this paper.
\item The link between the impedance criterion and the power-frequency transfer functions is analytically derived.
This is illustrated via the \ac{nfp} plots. 
\end{itemize}
One should note that the presented work does \emph{not} propose a new method of stability analysis, aiming to replace either state-space or impedance methods, but rather presents \emph{formal equivalencies} between existing methods.
The correctness of the proposed analytic developments are validated using a small microgrid test case, as well as a  transmission system benchmark, both implemented in Matlab/Simulink environment.

The remainder of this paper is organised as follows. 
Section~\ref{sec.state_space} presents the state-space representation and Section~\ref{sec.impedance} describes the derivation of the impedance matrices from the state-space representation, with and without the inclusion of the network fundamental frequency variations.
In Section~\ref{sec.frequency.link}, the link between frequency and apparent power is derived and expressed in terms compatible with impedance analysis.
The MG test case is presented in Section~\ref{sec.MG.test.case} while the large scale grid is studied in Section~\ref{sec.trans.case.study}.
Finally, a discussion regarding implementation aspects and challenges is found in Section~\ref{sec.discussion}, while the conclusion and suggestions for further research are presented in Section~\ref{sec.conclusion}.

\section{State-Space Representation}
\label{sec.state_space}
\begin{figure}[!t]
\centering
\includegraphics[width=0.42\columnwidth]{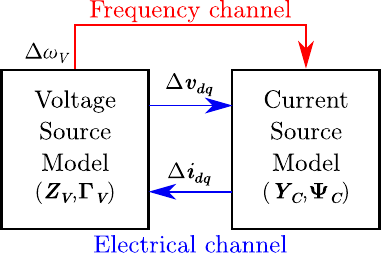}
\caption{Equivalent dynamic model of the system under study.} 
\label{fig.equivalent.model}
\end{figure}

For the notation used in the following, uppercase bold will signify matrices, lowercase bold will signify vectors and regular font will signify scalars. 
The general formulation of the linearised, state-space representation of a dynamic device is the following~\cite{ogata2010modern}:
\begin{equation}
\label{eq:linear_ode}
\begin{aligned}
\Delta \dot{\boldsymbol{x}}=\boldsymbol{A} \Delta \boldsymbol{x}+ \boldsymbol{B} \Delta \boldsymbol{u},
\;\; \Delta \boldsymbol{y}=\boldsymbol{C} \Delta \boldsymbol{x} + \boldsymbol{D} \Delta \boldsymbol{u},
\end{aligned}     
\end{equation}
where $\Delta$ stands for ``small perturbation'', $<\dot{}>$ is the derivative operator, $\boldsymbol{A}$, $\boldsymbol{B}$, $\boldsymbol{C}$, and $\boldsymbol{D}$ are the state, input, output and feed-forward matrices, respectively and $\boldsymbol{x}$, $\boldsymbol{u}$, and $\boldsymbol{y}$, are the state, input and output vectors, respectively. 
For simplicity, and without loss of generality, the outputs of the dynamic devices in this study are considered to be state variables, hence $\boldsymbol{D}=0$.
In~\eqref{eq:linear_ode}, all electrical variables, namely voltages ($u$) and currents ($i$), are commonly expressed in rotating $dq$ reference frames, as in: $\Delta \boldsymbol{v_{dq}}=[\Delta v_d \;\; \Delta v_q]^\intercal$, $\Delta \boldsymbol{i_{dq}}=[\Delta i_d \;\; \Delta i_q]^\intercal$, where $^{\intercal}$ is the transpose operator.
For the equivalency between the state-space representation and the impedance representation, two types of dynamic devices will be considered, as illustrated in Fig.~\ref{fig.equivalent.model}.
The first one is a \ac{vs} that sets the voltage at its connection bus and defines the frequency reference for all the other elements in the grid, while the second one is a \ac{cs} that sets the current injection at its connection bus.
It is noted that the spatial characteristics of the fundamental frequency are not considered in this study, with the fundamental frequency variation being treated as a global variable.

For the \ac{vs}, the inputs are the injected current $\Delta \boldsymbol{i_{dq}}$ at the connection bus, together with an arbitrary set of reference commands $\Delta \boldsymbol{u^*}$, while the outputs are the voltage at its connection point  $\Delta \boldsymbol{v_{dq}}$ and the frequency reference for the rest of the network $\Delta \omega_V$.
By substituting $\Delta \boldsymbol{u}=[\Delta \boldsymbol{i_{dq}} \; \; \Delta \boldsymbol{u^*}]^\intercal$ and $\Delta \boldsymbol{y}=[\Delta \boldsymbol{v_{dq}} \; \;  \Delta \omega_V]^\intercal$ in~\eqref{eq:linear_ode} and using the subscript $V$ to denote \ac{vs}, the state-space representation of the \ac{vs} becomes:
\begin{align}
\begin{split}
&\Delta \dot{
 \boldsymbol{x}}_{V}
= \boldsymbol{A_{V}}
\Delta \boldsymbol{x_{V}}
+
\boldsymbol{B_{i-V}} \Delta \boldsymbol{i_{dq}} 
+ \boldsymbol{B_{u^*-V}} \Delta \boldsymbol{u_V^*},
\\
&\begin{bmatrix}
\Delta \boldsymbol{v_{dq}} \\  \Delta \omega_V
\end{bmatrix}
= \boldsymbol{C_{V}} \Delta \boldsymbol{x}_{V}
,
\label{eq:vs_state_space}
\end{split}
\end{align} 
where the different subscripts of the input matrices $\boldsymbol{B}$ correspond to its respective input channel.

For the \ac{cs}, the inputs are the voltage at the connection bus $\Delta \boldsymbol{v_{dq}}$, the reference frequency $\Delta \omega_V$ set by the \ac{vs}, as well as some command signals $\Delta \boldsymbol{u^*}$, similar to the previous case.
The output is the current $\Delta \boldsymbol{i_{dq}}$ injected at the connection point.
Similarly to the case of the \ac{vs}, by substituting $\Delta \boldsymbol{u}=[\Delta \boldsymbol{v_{dq}} \; \; \Delta \boldsymbol{u^*} \; \;  \Delta \omega_V ]^\intercal$ and $\Delta \boldsymbol{y}=\Delta \boldsymbol{i_{dq}}$ in~\eqref{eq:linear_ode} and using the subscript $C$ to denote \ac{cs}, the state-space representation of the \ac{cs} becomes:
\begin{align}
\begin{split}
\Delta \dot{
\boldsymbol{x}}_{C}
&=
\boldsymbol{A_{C}} \Delta \boldsymbol{x}_{C}
+
\boldsymbol{B_{v-C}} \Delta \boldsymbol{v_{dq}} 
+ 
\boldsymbol{B_{u^*-C}} \Delta \boldsymbol{u_C^*}
+\boldsymbol{B_{\omega-C}} \Delta \omega_V ,\\
\Delta \boldsymbol{i_{dq}} 
&=
\boldsymbol{C_{C}} \Delta \boldsymbol{x}_{C}
,
\label{eq:cs_state_space}
\end{split}
\end{align} 
where the different subscripts of the input matrices $\boldsymbol{B}$ correspond to the respective input channels, similar to the previous case.
It should be noted that~\eqref{eq:vs_state_space} and~\eqref{eq:cs_state_space} (namely \ac{vs} and \ac{cs} representations) can be used to model the majority of the dynamic elements that are encountered in modern power systems, including \ac{gfm} and \ac{gfl} converters, synchronous generators and loads.
The specific derivation of the matrices in~\eqref{eq:vs_state_space} and~\eqref{eq:cs_state_space} from the modelling equations of \ac{gfm} and \ac{gfl} converters can be found in the literature~\cite{PogakuMicrogrid,AlbertoDeadTime}.
By combining~\eqref{eq:vs_state_space} and~\eqref{eq:cs_state_space}, the state-space model of the whole system, denoted with the subscript $SYS$, is derived as follows:
\begin{align}
\begin{split}
\overbrace{
\begin{bmatrix}
\Delta \dot{\boldsymbol{x}}_{V}
\\ 
\Delta \dot{\boldsymbol{x}}_{C}
\end{bmatrix}
}^{\Delta \dot{ \boldsymbol{x}}_{\boldsymbol{SYS}}}
&=
\overbrace{
\begin{bmatrix}
\boldsymbol{A_{V}}
&
\boldsymbol{B_{i-V}} \boldsymbol{C_{C}}
\\ 
\boldsymbol{B_{\omega-C}} \boldsymbol{C_{V}}
&
\boldsymbol{A_{C}}
\end{bmatrix}
}
^{\boldsymbol{A_{SYS}}}
\overbrace{
\begin{bmatrix}
\Delta \boldsymbol{x}_{V}
\\ 
\Delta \boldsymbol{x}_{C}
\end{bmatrix}
}
^{\Delta \boldsymbol{x_{SYS}}}
+
\boldsymbol{B_{u^*-V}} \Delta \boldsymbol{u_V^*}
+ 
\boldsymbol{B_{u^*-C}} \Delta \boldsymbol{u_C^*}
,
\\
\Delta \boldsymbol{y_{SYS}} 
&=
\boldsymbol{C_{SYS}} \begin{bmatrix}
\Delta \boldsymbol{x}_{V}
\\ 
\Delta \boldsymbol{x}_{C}
\end{bmatrix}
.
\label{eq:sys_state_space}
\end{split}
\end{align} 
\begin{figure}[!t]
\centering
\includegraphics[width=0.8\linewidth]{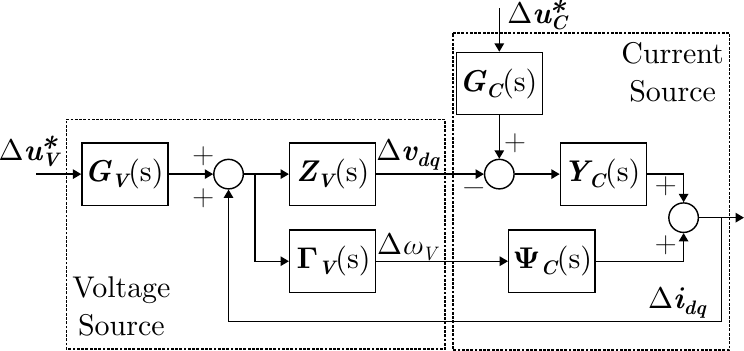}
\caption{System block diagram representation of the voltage and current source dynamic interaction.
}
\label{fig.block_diagram_detailed}
\end{figure}
It is noted that the output vector $\Delta \boldsymbol{y_{SYS}} 
$~(and hence also the matrix $\boldsymbol{C_{SYS}}$) can be arbitrarily selected depending on the states to be observed from the system.
For stability analysis based on the state-space method, the eigenvalues of $\boldsymbol{A_{SYS}}$ are examined, with the system being asymptotically stable when all eigenvalues have a negative real part~\cite{milano2020eigenvalue}.

One should note that all previous derivations assume a constant operating point.
For the specific case of the fundamental frequency operating point ($\omega_{V,o}$), its value depends on the effect of primary controllers (droops, etc.) and the corrections made by the secondary frequency controllers found in the system, if any.
\section{Impedance Derivation}
\label{sec.impedance}
In this section, the derivation of the impedance transfer functions from the state-space equations is detailed. 
First, the standard state-space and impedance relation is derived, without including the frequency dynamics of the \ac{vs}. 
This step can also be found in the literature, but is also included in this work for clarity and for the sake of completeness~\cite{amin2017small}.
Then the equivalency of the two representations for the purpose of stability analysis will be extended by also considering the frequency dynamics of the \ac{vs}, which is one of the main contributions of this work.
In the following derivations, operator $\Delta$ is dropped for the ease of notation and both matrices and vectors are denoted in uppercase (e.g., $\boldsymbol{V_{dq}}$ and $\boldsymbol{I_{dq}}$ for voltage and current Park vectors, respectively) to highlight the transition to the Laplace domain and for consistency with the common notation used in the literature.

\subsection{Equivalency Excluding Frequency Dynamics}
\label{sec.equivalency_nofreq}
First, the equivalency between the two representations is explained without considering the frequency dynamics.
For the \ac{vs} and~\eqref{eq:vs_state_space}, we consider the external command signals constant~(hence $\Delta\boldsymbol{u_V^*}=0$) and neglect the output frequency $\Delta \omega_V$, thus only focusing on the relationship between voltage and current.
Applying the Laplace transform in~\eqref{eq:vs_state_space}, and eliminating the internal states, the transfer function between the input current and the output voltage is obtained:
\begin{align}
\label{eq:impedance_V}
\boldsymbol{V_{dq}} (s)
&= 
\boldsymbol{Z_V}  (s)
\boldsymbol{I_{dq}}(s)
,    
\\
\label{eq:impedance_V_ss}
\boldsymbol{Z_{V}}(s)
&=
\boldsymbol{C_V}(s\boldsymbol{I}-\boldsymbol{A_V})^{-1}\boldsymbol{B_{i-V}},
\end{align}
where $s$ is the Laplace operator, $^{-1}$ is the inverse matrix operator and $\boldsymbol{I}$ is the unit matrix of suitable size.
Similarly, for the \ac{cs}, by neglecting the frequency input and the external commands ($\Delta\boldsymbol{u_C^*}= 0$, $\Delta \omega_V=0$), the transfer function between the input voltage and the output current is:
\begin{align}
\label{eq:admittance_I}
\boldsymbol{I_{dq}} (s)
&= 
\boldsymbol{Y_C}  (s)
\boldsymbol{V_{dq}}(s)
,    
\\
\label{eq:admittance_I_ss}
\boldsymbol{Y_{C}}(s)
&=
\boldsymbol{C_C}(s\boldsymbol{I}-\boldsymbol{A_C})^{-1}\boldsymbol{B_{v-C}}.
\end{align}
The symbols $\boldsymbol{Z_V}(s)$ and $\boldsymbol{Y_{C}}(s)$ represent the standard impedance and admittance transfer function matrices.
Their elements are scalar transfer functions, and their units are Ohms and Siemens, respectively, defined as:
\begin{equation}
\begin{aligned}
\boldsymbol{Z_V}(s)
=
\begin{bmatrix} 
Z_{V,dd}(s) & Z_{V,dq}(s) \\
Z_{V,qd}(s) & Z_{V,qq}(s) 
\end {bmatrix},\;\;
\;
\boldsymbol{Y_{C}}(s)
= 
\begin{bmatrix} 
Y_{C,dd}(s) & Y_{C,dq}(s) \\
Y_{C,qd}(s) & Y_{C,qq}(s) 
\end {bmatrix}. 
\end{aligned}
\end{equation}

Combining~\eqref{eq:impedance_V} and~\eqref{eq:admittance_I}, the block diagram of Fig.~\ref{fig.block_diagram_detailed} is derived. 
From it, and neglecting the frequency channel, the closed-loop characteristic equation of the system is:
\begin{equation}
\label{eq:closed_loop1}
\boldsymbol{H_1} (s) 
=
\boldsymbol{I}
+
\boldsymbol{Y_C}(s) \boldsymbol{Z_V}(s).
\end{equation}
Instead of the studying the closed-loop transfer function, the stability of the system can be inferred by applying the generalised Nyquist criterion to the open-loop~(minor loop gain)~\cite{sun2011impedance}.
For the system of Fig.~\ref{fig.block_diagram_detailed} and~\eqref{eq:closed_loop1}, the minor-loop is defined as:
\begin{equation}
\label{eq:open_loop1}
\boldsymbol{L_1} (s) 
=
\boldsymbol{Y_C}(s) \boldsymbol{Z_V}(s).
\end{equation}
Closed-loop stability can be guaranteed if the open-loop system satisfies the \ac{gnc}. 
This implies that each of the eigenvalues of $\boldsymbol{L_1}(s)$ should satisfy the Nyquist criterion for SISO systems.
In summary, when frequency variations are neglected, eigenvalues of $\boldsymbol{L_1}(s)$ fulfilling the Nyquist Criterion is equivalent to the eigenvalues of $\boldsymbol{A_{sys}}$~($\boldsymbol{B_{\omega-C}}$ being a zero matrix) having all negative real part.
The following section will detail how to lift this assumption, generalise and extend the equivalence of the two system representations by including network frequency variations.
\subsection{Equivalency Considering Frequency Dynamics}
\label{sec.equivalency_withfreq}
In the general case, the frequency variations $\Delta \omega_V$ from~\eqref{eq:cs_state_space} is non-zero. By applying the Laplace operator in~\eqref{eq:cs_state_space} and by using the superposition property of linear systems, the transfer function that relates the network frequency variation and the current injection of the \ac{cs} is calculated independently of the other system inputs.
By adding the contribution of the frequency variations, \eqref{eq:admittance_I} is modified as:
\begin{align}
\label{eq:admittance_I_extended}
{\boldsymbol{I_{dq}}}(s)
&=
{\boldsymbol{Y_C}} (s)\boldsymbol{V_{dq}}(s)
+ 
\boldsymbol{\Psi_C}(s) \Omega_V(s),
\\
\label{eq:psi_ss}
\boldsymbol{\Psi_{C}}(s)
&=
\boldsymbol{C_C} (s \boldsymbol{I} - \boldsymbol{A_{C}})^{-1}
\boldsymbol{B_{\omega-C}},
\end{align}
where 
\begin{equation}
\label{eq:psi_elements}
\boldsymbol{\Psi_{C}}(s)
=
\begin{bmatrix}
\Psi_{C,d}(s) \\  
\Psi_{C,q}(s)
\end{bmatrix}
\end{equation}
\noindent is called frequency-admittance vector and its units are A/(rad/s).
Similarly, by evaluating the last row in~\eqref{eq:vs_state_space}, the transfer function between the input current injection and the output frequency variation is calculated as:
\begin{align}
\label{eq:omega_v}
\Omega_V(s) 
&= 
\boldsymbol{\Gamma_V(s)} \boldsymbol{I_{dq}}(s), 
\\
\boldsymbol{\Gamma_V}(s) 
&=
\boldsymbol{C_C}(s \boldsymbol{I}-\boldsymbol{A_V})^{-1}\boldsymbol{B_{i-V}},
\label{eq:gamma_ss}
\end{align}
where 
\begin{equation}
\boldsymbol{\Gamma_{V}}(s)
=
[\Gamma_{V,d}(s) \; \Gamma_{V,q}(s)]
\end{equation}
is called the frequency-impedance vector and its units are (rad/s)/A.
Equation~\eqref{eq:omega_v} complements~\eqref{eq:impedance_V}, each defining one of the two outputs of the \ac{vs}.
It should be noted that the frequency-impedance and frequency-admittance are $1 \times 2$ and $2 \times 1$ vectors, respectively, hence having different dimensions compared with the standard impedance and admittance $2 \times 2$ matrices.
They represent concepts similar to the impedance and admittance in electrical circuits, yet adapted to represent the grid frequency dynamics.

By now considering the frequency variation channel in Fig.~\ref{fig.block_diagram_detailed}, the closed loop characteristic equation becomes:
\begin{equation}
\label{eq:closed_loop2}
\boldsymbol{H_2} (s) 
=
\boldsymbol{I}
+
\boldsymbol{Y_C}(s) \boldsymbol{Z_V}(s)
- 
\boldsymbol{\Psi_C}(s) 
\boldsymbol{\Gamma_V}(s)
.
\end{equation}
The equivalent open-loop transfer function matrix (i.e., minor loop) is:
\begin{equation}
\label{eq:open_loop2}
\boldsymbol{L_2}(s) 
= 
\boldsymbol{Y_C}(s) \boldsymbol{Z_V}(s)  
- 
\boldsymbol{\Psi_C}(s) \boldsymbol{\Gamma_V}(s).
\end{equation}
In summary Equations~\eqref{eq:admittance_I_extended}, \eqref{eq:omega_v} and \eqref{eq:open_loop2} detail how the known impedance representation is adapted to incorporate the frequency variation signal, extending the well-known equations \eqref{eq:admittance_I}, \eqref{eq:impedance_V} and \eqref{eq:open_loop1}, respectively.
Equations~\eqref{eq:psi_ss} and \eqref{eq:gamma_ss} shows the derivation of the frequency variation transfer functions from the state-space matrices, complementing \eqref{eq:impedance_V_ss} and \eqref{eq:admittance_I_ss} and completing the extension of the equivalency between the two representations for stability assessment with the inclusion of dynamic, network fundamental frequency variations.
The above are summarized and illustrated in Tables~\ref{table.impedance_extension} and~\ref{table.ss2tf}.
\begin{table}[!b]
\centering
\captionsetup{labelfont={color=black},textfont={color=black}}
\caption{Summary of impedance equation extensions}
\vspace{-0.1cm}
\arrayrulecolor{black}

\begin{tabular}{|l|l|l|}
\hline
 & Without frequency variation & With frequency variation \\
\hline
Current & \eqref{eq:admittance_I}: $\boldsymbol{I_{dq}} (s)
= 
\boldsymbol{Y_C}  (s)
\boldsymbol{V_{dq}}(s)$ 
&
\eqref{eq:admittance_I_extended}:
${\boldsymbol{I_{dq}}}(s)
=
{\boldsymbol{Y_C}} (s)\boldsymbol{V_{dq}}(s)
+ 
\boldsymbol{\Psi_C}(s) \Omega_V(s)$ \\
\hline
Voltage 
&
\eqref{eq:impedance_V}:
$\boldsymbol{V_{dq}} (s)
= 
\boldsymbol{Z_V}  (s)
\boldsymbol{I_{dq}}(s)$
&
\eqref{eq:impedance_V},\eqref{eq:omega_v}:
$\boldsymbol{V_{dq}} (s)
= 
\boldsymbol{Z_V}  (s)
\boldsymbol{I_{dq}}(s)$
,
$\Omega_V(s) 
= 
\boldsymbol{\Gamma_V(s)} \boldsymbol{I_{dq}}(s)$
\\
\hline
Minor loop
&
\eqref{eq:open_loop1}:
$\boldsymbol{Y_C}(s) \boldsymbol{Z_V}(s)$
&
\eqref{eq:open_loop2}:
$\boldsymbol{Y_C}(s) \boldsymbol{Z_V}(s)  
- 
\boldsymbol{\Psi_C}(s) \boldsymbol{\Gamma_V}(s)$ \\
\hline
\end{tabular}
\label{table.impedance_extension}
\end{table}
\begin{table}[!b]
\centering
\captionsetup{labelfont={color=black},textfont={color=black}}
\vspace{+0.2cm}
\caption{Summary of state-space to transfer function derivations}
\vspace{-0.1cm}
\arrayrulecolor{black}

\begin{tabular}{|l|l|}
\hline
Transfer function & Equation \\
\hline
Impedance
& 
\eqref{eq:impedance_V_ss}:
$\boldsymbol{Z_{V}}(s)
=
\boldsymbol{C_V}(s\boldsymbol{I}-\boldsymbol{A_V})^{-1}\boldsymbol{B_{i-V}}$
\\
\hline
Admittance
&
\eqref{eq:admittance_I_ss}: 
$\boldsymbol{Y_{C}}(s)
=
\boldsymbol{C_C}(s\boldsymbol{I}-\boldsymbol{A_C})^{-1}\boldsymbol{B_{v-C}}$
\\
\hline
Frequency-impedance 
& 
\eqref{eq:gamma_ss}:
$\boldsymbol{\Gamma_V}(s) 
=
\boldsymbol{C_C}(s \boldsymbol{I}-\boldsymbol{A_V})^{-1}\boldsymbol{B_{i-V}}$
\\
\hline
Frequency-admittance
& 
\eqref{eq:psi_ss}:
$\boldsymbol{\Psi_{C}}(s)
=
\boldsymbol{C_C} (s \boldsymbol{I} - \boldsymbol{A_{C}})^{-1}
\boldsymbol{B_{\omega-C}}$
\\
\hline
\end{tabular}
\label{table.ss2tf}
\end{table}
\section{Frequency to Apparent Power Dynamic Link}
\label{sec.frequency.link}
\subsection{The Network Frequency Perturbation Plot}
The \ac{nfp} plot is defined as the Bode plot of the transfer function that links the small-signal deviation of the fundamental frequency of the network $f_g$ with the active power output $P$ of a converter connected to the grid~\cite{mudalige2024control}:
\begin{equation}
\label{eq:nfp_def}
R(s)
=
\frac{\Delta P(s)}{\Delta f_g(s)}
=
2 \pi \frac{\Delta P(s)}{\Delta \omega_g(s)}
=
2 \pi \frac{\Delta P(s)}{\Omega_V(s)}
,
\end{equation}
with the fundamental grid frequency expressed using the notation of Section~\ref{sec.impedance}, i.e., $\Omega_V(s)=\mathcal{L}\{ \Delta \omega_V(t) \}=\mathcal{L}\{ \Delta \omega_g(t) \}$.
\subsection{Analytic Link between NFP and Frequency-Admittance}
\vspace{-0.1cm}

The active $P$ and reactive $Q$ power components injected by a shunt device (represented by the \ac{cs}) into the network (represented by the \ac{vs}) can be modelled using a power-invariant Park transform as follows:
\begin{equation}
\begin{aligned}
\label{eq.PQ}
P = v_d i_d + v_q i_q, \;\;
Q = v_q i_d -v_d i_q.
\end{aligned}
\end{equation}
By linearising~\eqref{eq.PQ}, its small-signal counterpart is derived:
\begin{equation}
\begin{aligned}
\label{eq.PQlinear}
\Delta P = V_{d,o} \Delta i_d
+I_{d,o} \Delta v_{d}
+V_{q,o} \Delta i_{q} 
+I_{q,o} \Delta v_{q},
\\
\Delta Q = V_{q,o} \Delta i_{d}
+I_{d,o} \Delta v_{q}
-V_{d,o} \Delta i_{q}
-I_{q,o} \Delta v_{d}
,
\end{aligned}
\end{equation}
where subscript ``o'' stands for ``operating point''.
By organising~\eqref{eq.PQlinear} in matrix form, its equivalent expression is derived:
\begin{equation}
\boldsymbol{S}
=
\begin{bmatrix}
\Delta P \\
\Delta Q
\end{bmatrix}
=
\underbrace{
\begin{bmatrix}
V_{d,o} & V_{q,o} \\
V_{q,o} & -V_{d,o}
\end{bmatrix}
}_{ \displaystyle \boldsymbol{V_o}}
\begin{bmatrix} 
\Delta i_d \\
\Delta i_q
\end{bmatrix}
+
\underbrace{
\begin{bmatrix}
I_{d,o} & I_{q,o} \\
-I_{q,o} & I_{d,o}
\end{bmatrix}
}_{ \displaystyle \boldsymbol{I_o}}
\begin{bmatrix} 
\Delta v_d \\
\Delta v_q
\end{bmatrix}.
\label{ec.apparent.power.derivation}
\end{equation}
By transforming in the Laplace domain, one gets:
\begin{equation}
\boldsymbol{S}(s) 
= 
\boldsymbol{V_o} \boldsymbol{I_{dq}}(s) 
+ 
\boldsymbol{I_o} \boldsymbol{V_{dq}}(s).
\label{ec.apparent.power.simple}
\end{equation}
The expression of the current in \eqref{eq:admittance_I_extended} can be replaced in \eqref{ec.apparent.power.simple} to find the full relation between the apparent power and the rest of the electrical signals involved in the power transfer between elements:
\begin{equation}
\boldsymbol{S}(s)
=
\boldsymbol{V_o}(\boldsymbol{Y_C}(s) \boldsymbol{V_{dq}}(s) + \boldsymbol{\Psi_C}(s) \Omega_V(s))
+
\boldsymbol{I_o} \boldsymbol{V_{dq}}(s).
\end{equation}
By reorganizing the latter expression in terms of voltage components and frequency:
\begin{equation}
\label{eq:nfp_freqadm_link}
\boldsymbol{S}(s)
=
\underbrace{
(\boldsymbol{U_o}\boldsymbol{Y_C}(s)+\boldsymbol{I_o})
}_{\displaystyle \boldsymbol{F_V}(s)}
\boldsymbol{V_{dq}}(s)
+
\underbrace{
\boldsymbol{V_o}\boldsymbol{\Psi_C}(s)
}_{\displaystyle \boldsymbol{F_\omega}(s)} \Omega_V(s),
\end{equation}
\noindent where $\boldsymbol{F_V}(s)$ and $\boldsymbol{F_\omega}(s)$ are the transfer functions that relate apparent power with voltage and frequency, respectively.
These functions have the following structure:
\begin{equation}
\label{eq:nfp_link_expressions}
\boldsymbol{F_V}(s)
=
\boldsymbol{U_o}\boldsymbol{Y_C}(s)+\boldsymbol{I_o}
=
\begin{bmatrix}
F_{{v_d,P}}(s) & F_{{v_q,P}}(s) \\
F_{{v_d,Q}}(s) & F_{{v_q,Q}}(s)
\end{bmatrix}
,
\; \;
\boldsymbol{F_{\omega}}(s)
=
\boldsymbol{V_o}\boldsymbol{\Psi_C}(s)
=
\begin{bmatrix}
F_{\omega P}(s) \\
F_{\omega Q} (s)
\end{bmatrix}, 
\end{equation}
where indices $v_d$, $v_q$ and $\omega$ stand for $dq$-axis voltage components and frequency, respectively, and $P$ and $Q$ stand for active and reactive power.
The elements of $\boldsymbol{F_{V}}(s)$ and $\boldsymbol{F_{\omega}}(s)$ relate the corresponding electrical variables of the network, i.e., voltage and frequency respectively, to the active and reactive power injections from the shunt generator devices (e.g., converters) and their detailed expressions can be found by expanding matrix expression~\eqref{eq:nfp_freqadm_link}.
In the following, for the economy of space, only the expression that is linked with the \ac{nfp} plot will be elaborated on. From \eqref{eq:nfp_freqadm_link} and \eqref{eq:nfp_link_expressions}, it can be easily identified that $F_{\omega P}(s)$ is a scaled version of the transfer function $R(s)$ from~\eqref{eq:nfp_def} that is used for the \ac{nfp} plot calculation. 
By evaluating~\eqref{eq:nfp_freqadm_link} for $F_{\omega P}(s)$ (using~\eqref{eq:psi_elements} and~\eqref{ec.apparent.power.derivation}), one gets:
\begin{equation}
\label{eq:Fwp}
F_{\omega P}(s)
= 
V_{d,o} \psi_{C,d}(s) 
+ 
V_{q,o} \psi_{C,q}(s),
\end{equation}
which precisely captures the analytic link between the \ac{nfp} transfer function and the frequency admittance.
In most situations, the equations of the device under study are expressed in a $dq$ rotating frame that is aligned with the $d$-axis component, hence: $V_{q,o}=0$. 
In this case:
\begin{equation}
\label{eq:nfp_freqadm_equiv}
F_{\omega P}(s)
= 
\frac{1}{2 \pi} R(s)
=
V_{d,o} \psi_{C,d}(s).
\end{equation}
Since in general $V_{d,o} \approx 1$ pu in normal conditions, \eqref{eq:nfp_freqadm_equiv} clearly demonstrates the equivalency between the \ac{nfp} transfer function and the frequency admittance quantity, thus bridging the gap between the two representations.

\begin{figure*}[!t]
\centering
\includegraphics[width=0.98\linewidth]{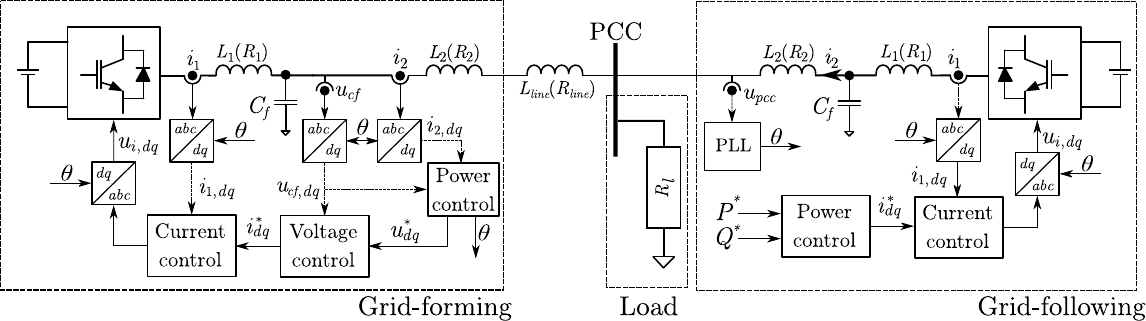}
\vspace{-0.1cm}
\caption{Electrical and control diagram of the test system.}
\vspace{-0.3cm}
\label{fig.diagrama.test.system}
\end{figure*} 
\section{Microgrid Case Study}
\label{sec.MG.test.case}
\subsection{Test System Description}
The analytical derivations presented in the previous sections are validated using a small microgrid system, shown in Fig.~\ref{fig.diagrama.test.system}.
This system consists of a \ac{gfm} converter, a \ac{gfl} converter and a load.
These devices are modelled as the \ac{vs} and the \ac{cs}, respectively.
This test configuration is selected as there is no stiff grid frequency regulation and the network fundamental frequency varies according to the \ac{gfm} converter control actions. 

\begin{table}[!b]
\centering
\vspace{-0.2cm}
\caption{Parameters of the \ac{gfm} converter.}
\vspace{-0.2cm}
\arrayrulecolor{black}
{
\begin{tabular}{|l|l|l|l|}
\hline
Parameter & Value & Parameter & Value\\
\hline
$L_{1}$ [mH] & 1.35 & $K_{P-CC}$ & 10.5\\ 
\hline
$L_{2}$ [mH] & 0.35 & $K_{I-CC}$ & 16000 \\
\hline
$C_f$ [$\mu$F] & 50 & $K_{P-VC}$ & 0.05\\
\hline
$R_{1}$ [m$\Omega$] & 100 & $K_{I-VC}$ & 390 \\
\hline
$R_{2}$ [m$\Omega$] & 30 & $K_{FF-VC}$ & 0.75\\
\hline
$m_p$ & $9.4 \cdot 10^{-5}$ & $n_q$ & 0.0013\\
\hline
$\omega_{c-P}$ [rad/s]& 31.41 & {$\omega_{c-Q}$ [rad/s]}& 31.41 \\
\hline
\end{tabular}
}
\label{table.GFM.parameters}
\vspace{-0.4cm}
\end{table}

\begin{table}[!b]
\centering
\vspace{+0.1cm}
\caption{Parameters of the \ac{gfl} converter.}
\vspace{-0.2cm}
\arrayrulecolor{black}
{
\begin{tabular}{|l|l|l|l|}
\hline
Parameter & Value & Parameter & Value\\
\hline
$L_{1}$ [mH] & 2.3 & $K_{P-CC}$ & 0.2351\\ 
\hline
$L_{2}$ [mH] & 0.93 & $K_{I-CC}$ & 24.349\\
\hline
$C_f$ [$\mu$F] & 8.8 & $K_{P-PLL}$ & 0.0143 \\
\hline
$R_{1}$ [m$\Omega$] & 72.2 & $K_{I-PLL}$ & 0.0519 \\
\hline
$R_{2}$ [m$\Omega$] & 29.2 &  &  \\
\hline
\end{tabular}
}
\label{table.GFL.parameters}
\end{table}

The \ac{gfm} device includes an internal current controller, an external voltage controller, and standard voltage and frequency droops~\cite{PogakuMicrogrid}.
The \ac{gfl} converter includes a current controller implemented in a \ac{srf}.
The \ac{srf} is synchronised with the $d$-axis component of \ac{pcc} voltage.
For this purpose, a \ac{pll} with the standard configuration is used~\cite{PLL_power_converter}.
The PLL of the GFL converter and the virtual swing equation equation of the GFM converter are modelled in full detail, meaning that the state-space model can accurately represent the frequency dynamics.
The state-space equations for the grid-forming and grid-following converters can be found in~\cite{PogakuMicrogrid,ARodriguez_Matrices}.
The linear matrices obtained during the linearisation process and the scripts used for merging them in a single state space model were uploaded to a public repository~\cite{ZENODO}, thus enhancing the reproducibility of the presented results. 
The parameters of the \ac{gfm} and the \ac{gfl} converters are shown in Tables~\ref{table.GFM.parameters} and \ref{table.GFL.parameters}, respectively, with the meaning of the electrical elements being shown in Fig.~\ref{fig.diagrama.test.system} and the dimensionless control gains having the usual meaning~\cite{PogakuMicrogrid}.

\subsection{Impedance-Based Modelling}
\begin{figure}[!t]
\centering
\includegraphics[width=0.7\linewidth]{
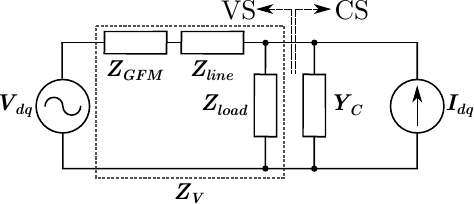}
\vspace{-0.2cm}
\caption{Small-signal representation of the system based on impedances.}
\label{fig.diagram.impedancia}
\end{figure}

The linear circuit used to represent the system of Fig.~\ref{fig.diagrama.test.system}  for the scope of impedance analysis is shown in Fig.~\ref{fig.diagram.impedancia}. 
In this figure, $\boldsymbol{Z_{GFM}}(s)$ is the impedance of the \ac{gfm} converter, $\boldsymbol{Z_{line}}(s)$ is the line impedance, $\boldsymbol{Z_{load}}(s)$ is the impedance of the load, $\boldsymbol{Z_{V}}(s)$ is the impedance of the \ac{gfm} converter plus the line and the load (i.e., the components that constitute the system \ac{vs}), and $\boldsymbol{Y_{C}}(s)$ is the admittance of the \ac{gfl} converter (i.e., the components that constitute the system \ac{cs}).
The values of these impedances/admittances are obtained using measurements, by applying the procedure described in~\cite{Small-Signal_Stability_Analysis_Based_Measured}.
The identification method is performed by using a frequency sampling method.
Specifically, for each frequency sampling point, two tones are injected to identify the impedance/admittance, and one to identify the grid frequency dynamics.
The tones are demodulated using a single-frequency demodulation technique (see~\cite{Lyons2004}, Chapter 8), followed by a high-order low-pass filter.
This method, despite being slower than the FFT or other demodulation techniques, ensures high selectivity and robustness against noise. 
The \ac{gnc} is then used to analyse the stability limits of the interconnected devices.
\vspace{-0.2cm}

\subsection{Comparison between Impedance and State-space Models}
\begin{figure}[!t]
\centering
\includegraphics[width=0.5\columnwidth]{
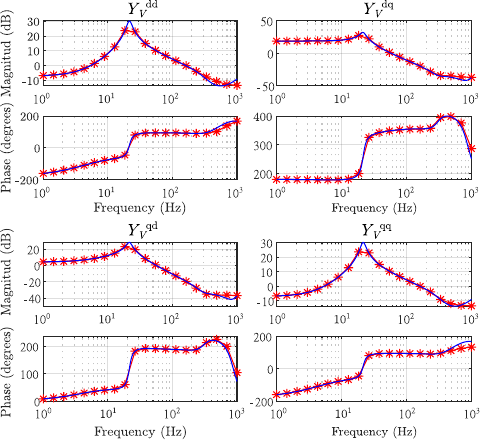}
\vspace{-0.1cm}
\caption{\ac{gfm} converter admittance. (blue) Theoretical, and (red) identified.}
\label{fig.GFM_Admittance.Dynamics}
\end{figure}

\begin{figure}[!t]
\centering
\includegraphics[width=0.5\columnwidth]{
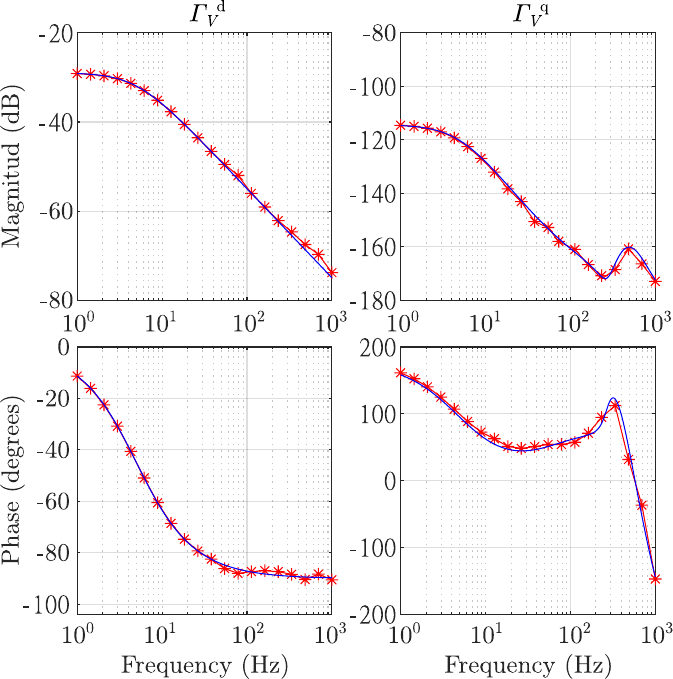}
\vspace{-0.1cm}
\caption{\ac{gfm} converter frequency impedance. (blue) Theoretical, and (red) identified.}
\label{fig.GFM_Frequency.Dynamics}
\end{figure}

\begin{figure}[!t]
\centering
\includegraphics[width=0.5\columnwidth]{
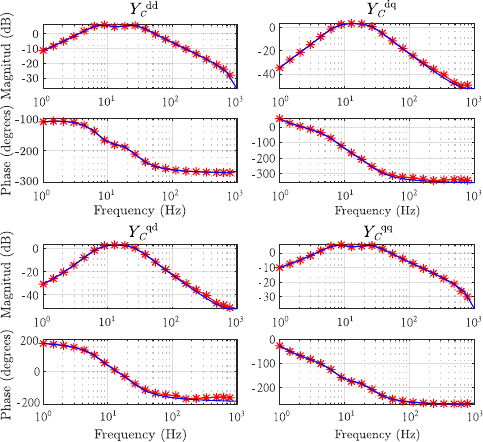}
\vspace{-0.2cm}
\caption{\ac{gfl} converter admittance. (blue) Theoretical, and (red) identified.}
\label{fig.GFL_Admittance.Dynamics}
\end{figure}

\begin{figure}[!t]
\centering
\includegraphics[width=0.5\columnwidth]{
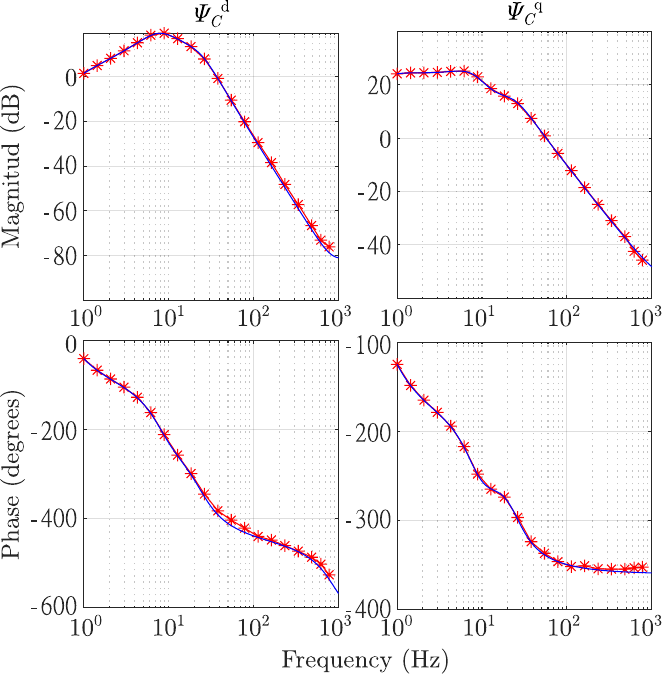}
\vspace{-0.1cm}
\caption{\ac{gfl} converter frequency admittance. (blue) Theoretical, and (red) identified.}
\label{fig.GFL_Frequency.Dynamics}
\end{figure}

For both cases, the results obtained using the impedance methodology will be compared to those obtained using the state-space methodology.
The analytical models used for the comparison of this section can be found in~\cite{ZENODO}. 
First, Fig.~\ref{fig.GFM_Admittance.Dynamics} and Fig.~\ref{fig.GFM_Frequency.Dynamics} show the response in the frequency domain of the admittance and the frequency-admittance of the \ac{vs}, respectively.
The results for the state-space method are presented in blue, and the results for the impedance method are presented in red.
It can be seen that in both cases the modelling approaches lead to similar results.
The results show an excellent match below 100~Hz, while some small differences can be observed in the range of 100~Hz to 1~kHz.
These small errors come from several sources. 
First, the gain of the plant at those frequencies is relatively low and, therefore, a large disturbance should be injected to measure the admittance characteristic.
This signifies that it becomes increasingly difficult to obtain accurate results while maintaining a reasonable disturbance amplitude in high frequencies.
Also, as the models were built using Matlab/Simulink and its SimPowerSystems toolbox, the impedance measurements were obtained using numerical methods, in which the time-step of simulations cannot be increased indefinitely.
All these elements decrease the precision of the results.
However, it should be noted that the focus of the paper are the low-frequency dynamics, and the system is far away from instability at high frequency. 
Fig.~\ref{fig.GFL_Admittance.Dynamics} and Fig.~\ref{fig.GFL_Frequency.Dynamics} show the admittance and the frequency admittance of the \ac{cs}.
The results follow the same trend as those of the \ac{vs}.

\subsection{Impact of Grid Frequency Dynamics on the GNC}
The impact of the grid frequency dynamics on the stability of the interconnected system is assessed with the eigenvalues of the minor loop $\boldsymbol{L}(s)$ for $s=j \omega $, with $\omega \in (-\infty, \infty) $.
Since $\boldsymbol{L}(j \omega)$ is a square matrix of second order, it has two eigenvalues: $\lambda_1(j \omega)$ and $\lambda_2(j \omega)$.
These eigenvalues will be depicted in polar coordinates so that the \ac{gnc} can be applied.
Due to their illustrative properties, Bode diagrams will be used to showcase the effect of the grid frequency dynamics at specific frequencies.
In the following expressions, if the grid frequency dynamics is not considered (i.e., $\boldsymbol{L_1} (s) =\boldsymbol{L}'(s) = \boldsymbol{Y_C}(s) \boldsymbol{Z_V}(s)$), then the eigenvalues will be denoted as $\lambda_1'(j \omega)$ and $\lambda_2'(j \omega)$.
In addition, the frequency dependence of the eigenvalues will be dropped for simplicity.


\begin{figure}[!t]
\centering
\includegraphics[width=0.6\columnwidth]{
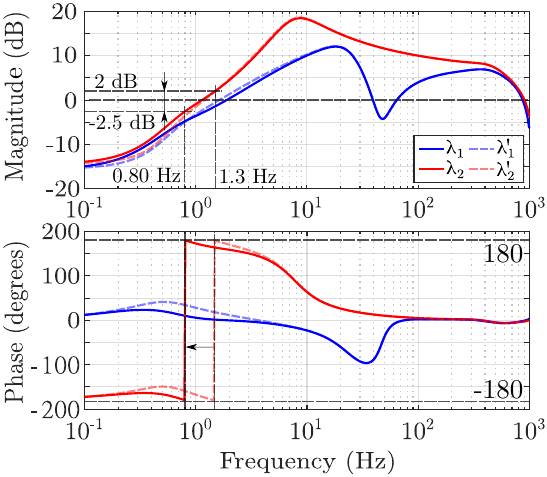}
\caption{Bode diagram of $\lambda_1$, $\lambda_1'$, $\lambda_2$ and $\lambda_2'$, for the studied system.}
\label{fig.new.vs.old.imp_freqDomain}
\end{figure}

Fig.~\ref{fig.new.vs.old.imp_freqDomain} shows the Bode diagram of the system eigenvalues, with and without the contribution of the grid frequency dynamics.
Despite the values being relatively similar, an important mismatch in the phase can be observed at frequencies around 1~Hz.
The stability of the system is assessed by checking if the magnitude is above (unstable) or below (stable) $0$~dB for the frequency when the phase crosses $-180$ degrees.
The phase of the first eigenvalue never crosses $-180$ degrees.
This means that this eigenvalue does not lead to instability and can be neglected.
For the case of the second eigenvalue, the phase is shifted to lower frequencies when the grid frequency dynamics are considered.
This fact leads to relevant changes in the system stability, as the magnitude for which the phase crosses $-180$ degrees is significantly modified.
Due to this modification, the system stability assessment changes from unstable to stable.
Therefore, the Bode diagram is considered of interest for finding the frequency in which stability is compromised.

\begin{figure}[!t]
\centering
\includegraphics[width=0.6\columnwidth]{
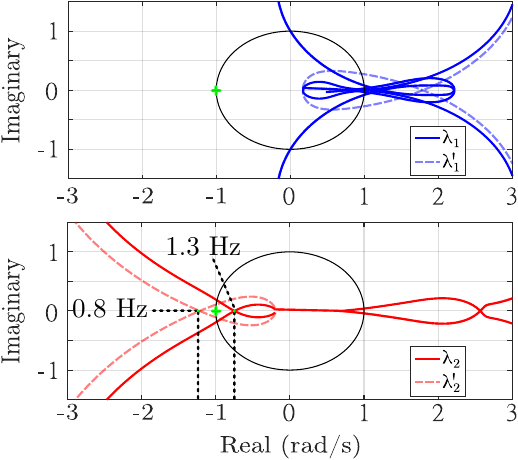}
\caption{Nyquist plot of the eigenvalues of the minor loop, with and without the frequency dynamics.}
\label{fig.new.vs.old.imp_Nyquist}
\end{figure}

The implications of these differences can be further highlighted using the Nyquist plot.
Fig.~\ref{fig.new.vs.old.imp_Nyquist} shows the Nyquist plot for the two eigenvalues.
The unit circle, denoting the critical point $(-1,0)$, is marked with black.
It can be observed that the effects of the grid frequency dynamics become clearly visible in the Nyquist plots.
For $\lambda_1$, despite the differences, these do not affect the stability of the system, as the phase is always far from $-1$.
For the case of $\lambda_2$, there are also important differences.
For high frequencies, $\lambda_2$ and $\lambda'_2$ do not differ.
However, at low frequencies, the crossing of $-1$ is significantly different.
Indeed, on the one hand, if the frequency dynamics of the grid is taken into account ($\lambda_2$), the point $-1$ is not encircled and the system is (theoretically) stable.
On the other hand, if the frequency dynamics of the grid is not considered ($\lambda'_2$), the \ac{gnc} predicts that the system will be unstable.
From these results, it can be concluded for this particular case study that the inclusion of the grid frequency dynamics is of paramount importance.
Additionally, it can be concluded that:
\begin{enumerate}
\item In this case, stability in the low-frequency range is mainly linked to one eigenvalue.
\item The \ac{gnc} without frequency dynamics yields an erroneous stability assessment.
\end{enumerate}

It should be noted that the real system does not oscillate, as all system modes are relatively well damped. 
However, the parameters of the test system can be easily modified in order to have oscillatory responses (e.g., by changing the bandwidth of the current controller or the PLL) at different frequencies, both at fast and slow timescales. 


\vspace{-0.2cm}

\subsection{Validation of Power Relationships}
\begin{figure}[!t]
\centering
\includegraphics[width=0.55\columnwidth]{
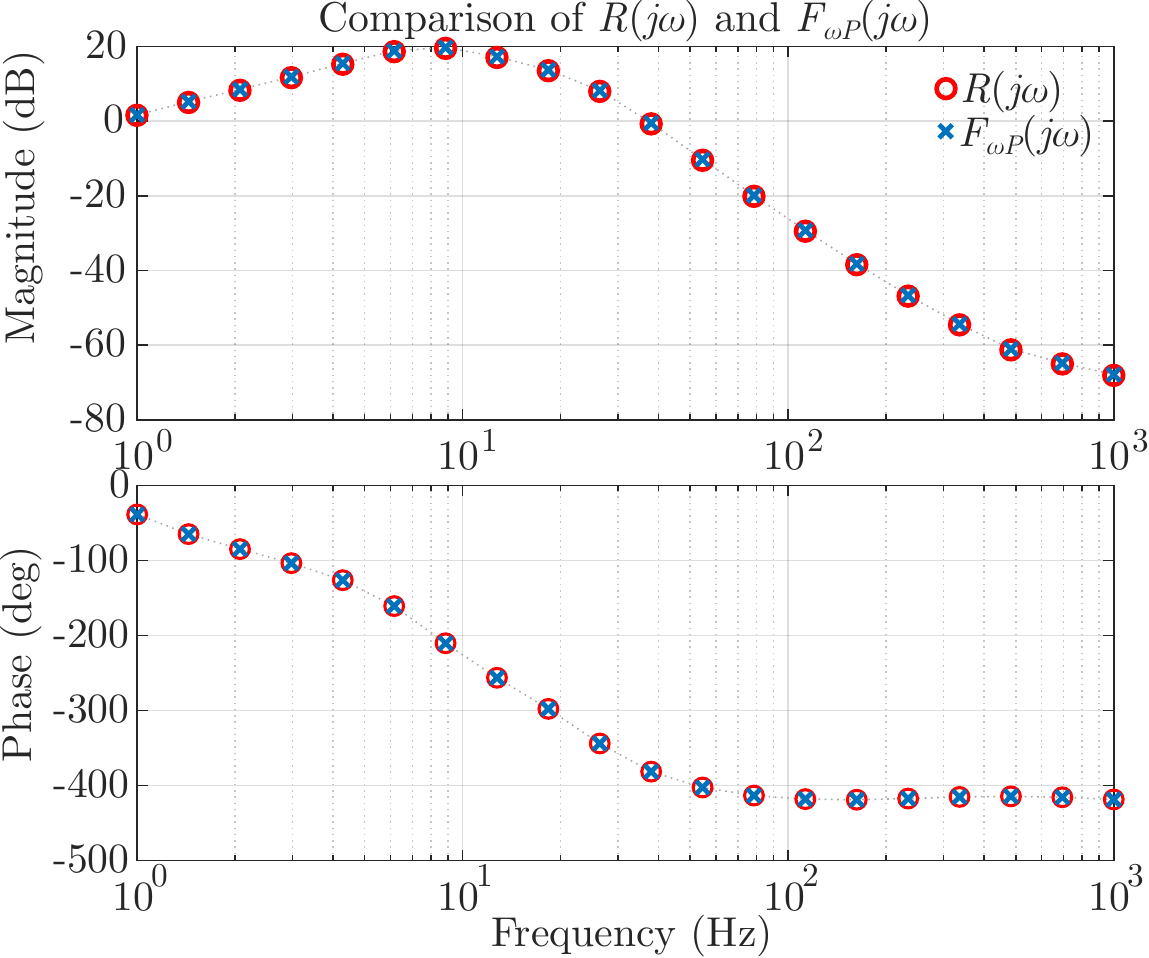}
\caption{Comparison of the NFP for the microgrid case study, depicted as $R(j \omega)$, and $F_{\omega P}(j \omega)$.}
\label{num.result1.validation.power.freq}
\end{figure}

Fig.~\ref{num.result1.validation.power.freq} shows measurements of the \ac{nfp} and $F_{\omega P}(j \omega)$, for the \ac{gfl} converter.
For this case, the bandwidth of the current controller was 100~Hz while the bandwidth of the \ac{pll} was 1~Hz.
The two curves perfectly coincide, validating the expression derived in (\ref{eq:nfp_freqadm_equiv}).

\subsubsection{Simulation Parameters}
\label{sec.parameters.test}
To validate the theoretical analyses, numerical simulations have been performed.
For that purpose, a non-linear model was developed using Matlab/Simulink and its SimPowerSystems toolbox.
The nominal power, voltage, and frequency of the system were set to 15~KW, 380~V and 50~Hz, respectively.
The rest of the hardware and control parameters can be found in Tables \ref{table.GFM.parameters} and \ref{table.GFL.parameters}.
The results presented in the following subsection confirm the previously presented stability predictions.

\subsubsection{Validation of Frequency Dynamics Contribution}
\begin{figure}[!t]
\centering
\includegraphics[width=0.55\columnwidth]{
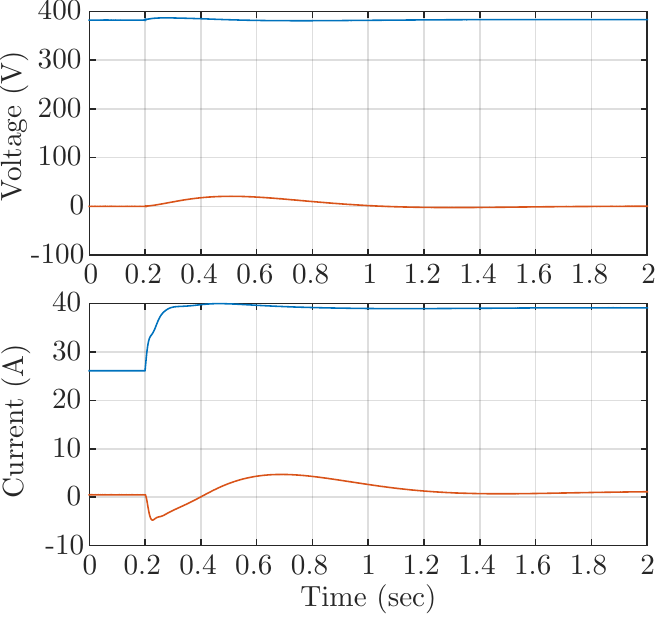}
\caption{Result validating closed-loop stability. 
(top) PCC voltage and (bottom) \ac{gfl} converter current, for a 5~kW step in the active power reference. (blue) $d$ and (red) $q$ axis component.}
\label{num.result1.validation.freq}
\end{figure}

Fig.~\ref{num.result1.validation.freq} shows a numerical simulation result of the test system, for a 5~kW step change in the active power set point of the \ac{gfl} converter (starting at 10~kW).
It can be seen that the system is stable, as predicted by the \ac{gnc} with the frequency dynamics included, shown in Fig.~\ref{fig.new.vs.old.imp_Nyquist}.
This result confirms that it is important to consider the grid frequency dynamics for an adequate stability assessment, as the impedance criterion without frequency dynamics, also shown in Fig.~\ref{fig.new.vs.old.imp_Nyquist}, predicted an unstable operation for the same case.
One should note that the FFT has not been used to analyse the time domain responses, since its application to non-stationary signals may lead to inaccurate results. 
However, the FFT could be helpful to analyse real, noisy measurements and obtain its spectra.

\section{Transmission System Case Study}
\label{sec.trans.case.study}
\begin{figure}[!b]
\centering
\includegraphics[width=0.65\columnwidth]{
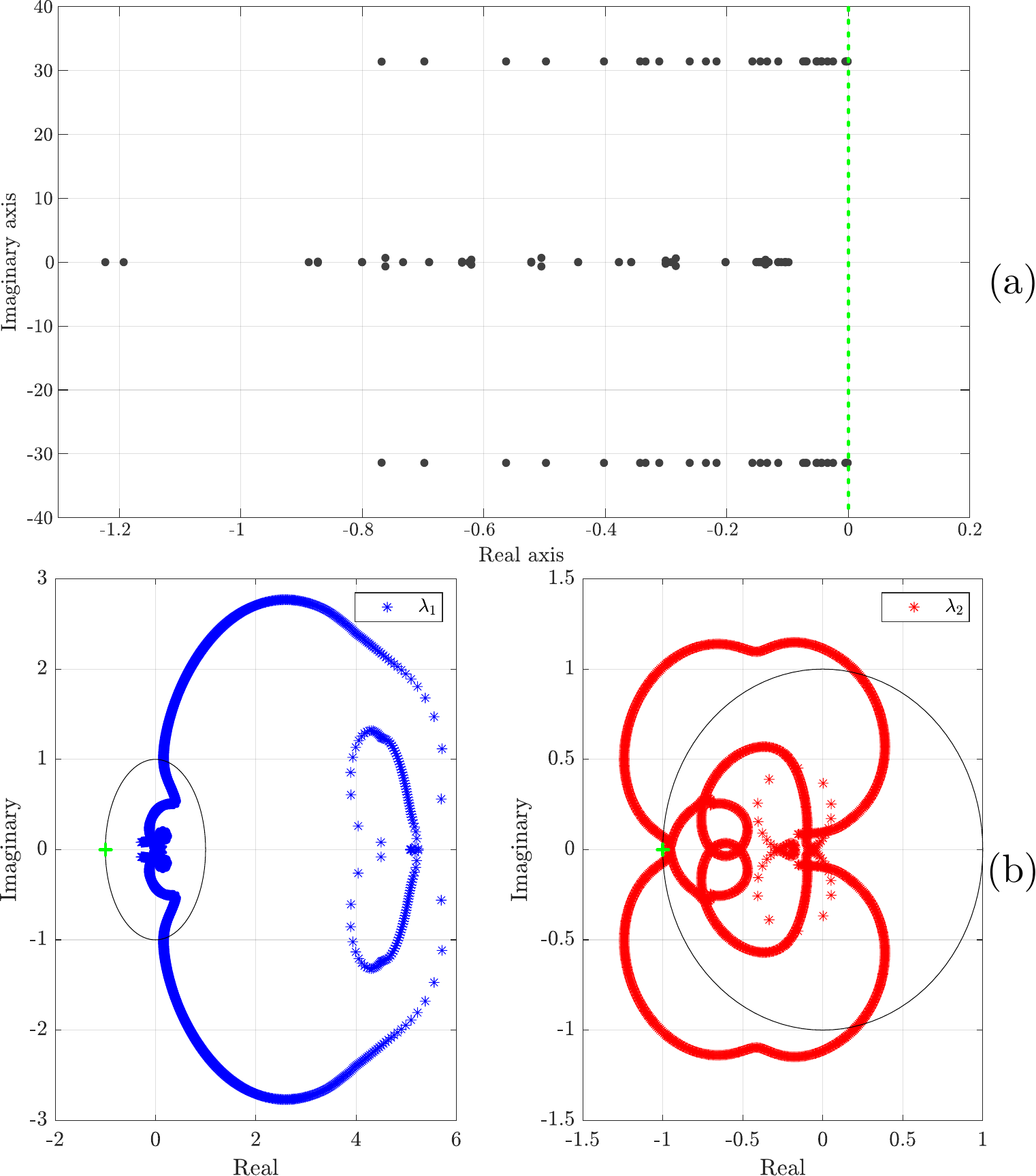}
\caption{Stability analysis results for the New England benchmark model. (a) Eigenvalues from the full state-space representation and (b) Nyquist plots from the eigenvalues of the \ac{gnc} with fundamental frequency dynamics included.}
\label{fig.new_england_eigen_nyquist}
\end{figure}
\begin{figure}[!t]
\centering
\includegraphics[width=0.65\columnwidth]{
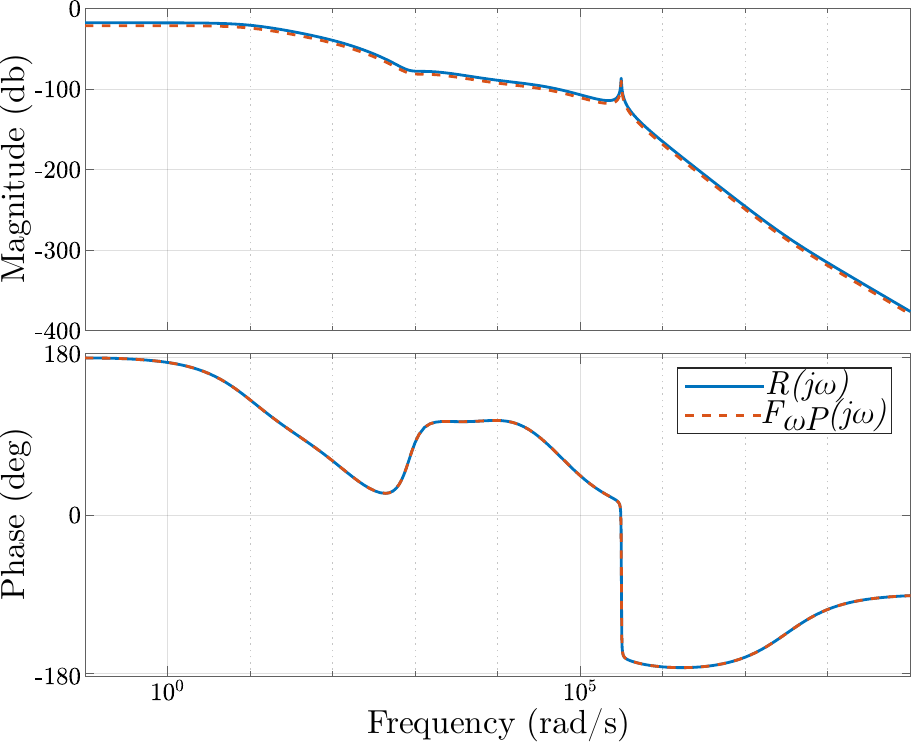}
\caption{Comparison of the NFP for the transmission system case, depicted as $R(j \omega)$, and $F_{\omega P}(j \omega)$.}
\label{fig.new_england_bode}
\end{figure}
In order to validate the theoretical contributions on a large power system, a case study was implemented based on a modified version of the well-known, New England transmission system.
In it, 35\% percent of the original synchronous machines were replaced by converter-based generators, operating in both \ac{gfm} and \ac{gfl} control modes.
The remaining synchronous machines were modelled using standard 6-th order models and were equipped with appropriate voltage and speed regulators~\cite{kundur1994power}.
The \ac{gfm} and \ac{gfl} control structures that were used for the converters were similar to the ones depicted in Fig.~\ref{fig.diagrama.test.system}.
Their detailed control schemes can be found in~\cite{11224594}.
The parameters that were used for all the participating devices can be found in~\cite{STAMP_Github}.
The derivation of the linear models for all the individual devices and the linear model of the complete system was performed using STAMP, a Matlab-based toolbox for automatic small-signal modelling~\cite{arevalo2025matlab}.
STAMP expedites the linear modelling process, achieving the calculation of the case study linear model, along with its eigenvalues, in 93 seconds in a conventional laptop, equipped with a 13th Gen Intel(R) Core(TM) i7-1355U (1.70 GHz) processor and a 40 GB RAM. 

In order to apply the \ac{gnc}, the complete system was partitioned as follows.
The \ac{gfl} converter connected at bus 14 was considered as the \ac{cs}, while the rest of the system was considered as the \ac{vs}.
Fig.~\ref{fig.new_england_eigen_nyquist} shows the stability analysis results for the New England transmission system case study.
They include the eigenvalues of the state matrix, resulting from the full state-space representation of the complete system (found in~\eqref{eq:sys_state_space}), as well as the Nyquist plots of the eigenvalues, resulting from the \ac{gnc} with fundamental frequency dynamics included (found in~\eqref{eq:open_loop2}).
It can be seen that both methods agree in the stability assessment of the network, indicating stable operation, with all eigenvalues being on the left-hand semi-plane and the Nyquist plots not encircling the critical point $(-1,0)$.
It can also be seen that despite the stable operation, the stability margin is not large, with many eigenvalues lying close to the imaginary axis and with the Nyquist plot of eigenvalue $\lambda_2$ being close to the critical point.
Fig.~\ref{fig.new_england_bode} shows a comparison on the system level between the \ac{nfp} transfer function, and the one calculated from \eqref{eq:Fwp}.
Both curves match perfectly, validating the correctness of the mathematical formulation.
\section{Discussion}
\label{sec.discussion}
Several practical implementation challenges and limitations exist for both state-space and transfer function-based approaches for stability analysis.
\subsection{State-Space Methods}
For state-space, eigenvalue analysis, analytical approaches are often preferred.
This requires significant efforts in terms of deriving the linear models, especially when considering large-scale systems.
Currently, automatized software toolboxes are available, which significantly expedite the small-signal modelling process~\cite{arevalo2025matlab,tomas2025vector}. 
Lowering the computational requirements for the linear analysis of large-scale power systems with high integration of power converters, along with the eigenvalue tracking across different system parameter sweeps, is currently an open research problem~\cite{bouterakos2025eigenvalue}.
Another practical challenge for eigenvalue-based stability analysis is the integration of black-box models which protect the intellectual property of converter manufacturers~\cite{smith2024black,Bollerslev2025RootCause}.
Currently, there is no universally agreed format for black-box models oriented for eigenvalue analysis, while the impact of said format to the analysis accuracy remains unclear~\cite{arbogast2025impact}.
\subsection{Transfer Function Methods}
For scenarios where detailed models are not available, estimated transfer function techniques (such as the impedance and power-frequency analyses) are typically preferred.
However, these techniques also face design and implementation challenges, such as choosing the appropriate type and magnitude for the perturbation signals, measured variables filtering and synchronization, etc.~\cite{mohammed2022fast}.
This is particularly challenging in the very-low frequency range, where seconds (or even minutes) are needed to measure one point of the frequency response~\cite{moutevelis2024virtual}. 
For multi-converter systems, the non-trivial task of estimating the transfer functions of all connected converters is necessary~\cite{sun2022frequency}.
These challenges are exacerbated when considering that the estimated transfer functions are dependent on the grid conditions (weak or strong), on the operating point of the system, as well as on the network topology, which are all subject to variations~\cite{benedetti2025exploring,desai2026effect}.
This necessitates the re-estimation of transfer function periodically, leading to increased computational requirements.
To address these challenges, current research is focused both on efficient algorithms for rapid impedance transfer function estimation~\cite{zhou2024rapid,mohammed2022fast,haberle2023mimo}, as well as on the integration of these algorithms in automatized toolboxes~\cite{garcia2026siad,garcia2026z}.
Combining the two approaches, a new analysis paradigm can be envisioned, where both analytic and estimation-based approaches coexist~\cite{11328961}.
This work contributes in this direction, providing a formal framework for the equivalency of the different methods currently used in the literature for power system stability studies.

\section{Conclusion}
\label{sec.conclusion}
In this work, the equivalency of three alternatives for the modelling of the fundamental frequency dynamic variations in power systems for small-signal stability studies have been presented.
These three methods include the state-space, the power-frequency, illustrated via the \ac{nfp} plots, and the impedance representations.
The latter has been used as a link to present closed-form expressions that relates all three methods in an analytic way.
Specifically, the recently proposed transfer functions that relate the fundamental frequency of the grid to the electric variables of the system were shown to be derived from a generic state-space formulation of the power system.
Additionally, under specific conditions, they were shown to be proportional to the transfer function that connects the active power injection of generator devices and the fundamental frequency disturbance, which are currently used for the \ac{nfp} plots in power converter grid-code compliance test applications.
A microgrid test system consisting of one \ac{gfm} and one \ac{gfl} converter has been used as a benchmark system, as it has pronounced variations in the system frequency.
Detailed numerical simulations were performed to validate the main findings.
An additional case study based on a modified version of the well-known New England transmission system with increased converter-based generation was also implemented to validate the consistency of the theoretical results in a large system.

The results obtained confirmed that the impedance modelling and the state-space modelling produce identical responses if the grid frequency dynamics is included in the impedance model.
Only slight differences were observed at high frequencies (around 1 kHz), due to the method used for the impedance estimation.
However, if the grid frequency dynamics is not taken into account, some important mismatches can be produced, especially in the phase at low frequencies (around 1~Hz).
These errors in the impedance lead to an inaccurate stability assessment and emphasise the importance of taking into account the grid frequency dynamics. 
In addition to that, if the impedance criterion includes the grid frequency dynamics, more conservative assessment results are produced compared to the case where the grid frequency dynamics are not considered.
Regarding the relationship between active power and the frequency-admittance, it has been shown that the two transfer functions, the one defined by the \ac{nfp} and the one calculated from the extended impedance criterion with included grid frequency dynamics, match precisely for the considered case study.
No inconsistencies between the different stability analysis methods and the produced theoretical expressions were observed for the large system.

In future work, it would be of interest to consider more sophisticated identification methods to calculate impedance and frequency-impedance transfer functions, thus alleviating the mismatches observed in the high-frequency range. 
In addition, it would be of interest to further apply the frequency-impedance concept for power system estimation, e.g., for the characterization of power system inertia and control applications.
The detailed analysis of test systems based on synchronous generators using the impedance criterion with frequency dynamics could also be of interest, as could also the study of the effect of the frequency spatial characteristics on the dynamic characterization of the system.
Finally, the analytical expressions for the impedance criterion in alternative coordinate systems, e.g., in terms of complex frequency variables~\cite{milano2021complex,moutevelis2023taxonomy}, will also be sought. 

\begingroup
\footnotesize
\setlength{\parskip}{0.5pt}
\begin{spacing}{1}
\setstretch{0.5}
\bibliography{Bibliografia}

@book{Lyons2004,
title={Understanding digital signal processing, 3/E},
author={Lyons, Richard G},
year={2004},
publisher={Pearson Education India}}

@INPROCEEDINGS{ARodriguez_Matrices,
author={Rodríguez-Cabero, Alberto and Prodanovic, Milan},
booktitle={IECON 2016 - 42nd Annual Conference of the IEEE Industrial Electronics Society}, 
title={{Stability Analysis for Weak Grids with Power Electronics Interfaces}}, 
year={2016},
volume={},
number={},
pages={2402--2407},
}

@misc{ZENODO,
  author = {Pablo Rodriguez-Ortega and Dionysios Moutevelis and Javier Roldan-Perez and Milan Prodanovic},
  title = {{State-Space Models of Grid-Forming and Grid-Following Converters}},
  doi = {10.5281/zenodo.18364144},
  year = {2026},
  howpublished = {\url{https://zenodo.org/records/18364145}},
}

@article{SIMON2022108528,
title = {Extraction of Frequency-Dependent Impedances of Residential Loads in Low-Voltage Grids for Harmonic Stability Assessment},
journal = {Electric Power Systems Research},
volume = {212},
pages = {108528},
year = {2022},
author = {Sandor Simon and Alexander Winkens and Antonello Monti and Andreas Ulbig}
}

@ARTICLE{AlbertoDeadTime,
author={A. {Rodr\'{\i}guez-Cabero} and M. {Prodanovic} and J. {Rold\'{a}n-Perez}},
journal={IEEE Transactions on Sustainable Energy}, 
title={{Analysis of Dynamic Properties of VSCs Connected to Weak Grids Including the Effects of Dead Time and Time Delays}}, 
year={2019},
volume={10},
number={3},
pages={1066--1075},
}

@book{kundur1994power,
title={Power System Stability and Control},
author={Kundur, P. and Balu, N.J. and Lauby, M.G.},
series={EPRI power system engineering series},
year={1994},
publisher={McGraw-Hill}
}

@ARTICLE{PogakuMicrogrid, 
author={N. Pogaku and M. Prodanovic and T. C. Green}, 
journal={IEEE Transactions on Power Electronics}, 
title={{Modeling, Analysis and Testing of Autonomous Operation of an Inverter-Based Microgrid}}, 
year={2007}, 
volume={22}, 
number={2}, 
pages={613--625},  
month={March}
}

@ARTICLE{Nyquist_Eigen_Analisys,
author={Fan, Lingling and Miao, Zhixin},
journal={IEEE Transactions on Power Systems}, 
title={{Admittance-Based Stability Analysis: Bode Plots, Nyquist Diagrams or Eigenvalue Analysis?}}, 
year={2020},
volume={35},
number={4},
pages={3312--3315}
}

@ARTICLE{Boroyevich,
author={Wang, Shike and Liu, Zeng and Liu, Jinjun and Boroyevich, Dushan and Burgos, Rolando},
journal={IEEE Transactions on Power Electronics}, 
title={{Small-Signal Modeling and Stability Prediction of Parallel Droop-Controlled Inverters Based on Terminal Characteristics of Individual Inverters}}, 
year={2020},
volume={35},
number={1},
pages={1045-1063}
}

@article{gong2020dq,
title={{DQ-Frame Impedance Measurement of Three-Phase Converters Using Time-Domain MIMO Parametric Identification}},
author={Gong, Hong and Wang, Xiongfei and Yang, Dongsheng},
journal={IEEE Transactions on Power Electronics},
volume={36},
number={2},
pages={2131--2142},
year={2020},
publisher={IEEE}
}

@article{luhtala2018implementation,
title={{Implementation of Real-Time Impedance-Based Stability Assessment of Grid-Connected Systems Using MIMO-Identification Techniques}},
author={Luhtala, Roni and Roinila, Tomi and Messo, Tuomas},
journal={IEEE Transactions on Industry Applications},
volume={54},
number={5},
pages={5054--5063},
year={2018},
publisher={IEEE}
}

@ARTICLE{Small-signal_Methods_Review,
author={Sun, Jian},
journal={IEEE Transactions on Power Electronics}, 
title={{Small-Signal Methods for AC Distributed Power Systems–A Review}},
year={2009},
volume={24},
number={11},
pages={2545--2554}
}

@ARTICLE{Review_Small_signal_modelling,
author={Yue, Xiaolong and Wang, Xiongfei and Blaabjerg, Frede},
journal={IEEE Transactions on Power Electronics}, 
title={{Review of Small-Signal Modeling Methods Including Frequency-Coupling Dynamics of Power Converters}}, 
year={2019},
volume={34},
number={4},
pages={3313-3328}
}

@ARTICLE{Impedance-Based_Whole_System_Modelling,
author={Gu, Yunjie and Li, Yitong and Zhu, Yue and Green, Timothy C.},
journal={IEEE Transactions on Power Systems}, 
title={{Impedance-Based Whole-System Modeling for a Composite Grid via Embedding of Frame Dynamics}}, 
year={2021},
volume={36},
number={1},
pages={336-345},
}

@ARTICLE{Power_electronic_integration,
author={Blaabjerg, F. and Chen, Zhe and Kjaer, S.B.},
journal={IEEE Transactions on Power Electronics}, 
title={{Power Electronics as Efficient Interface in Dispersed Power Generation Systems}}, 
year={2004},
volume={19},
number={5},
pages={1184--1194}
}

@ARTICLE{Impedance-based_Grid_Synchronization,
author={Wen, Bo and Dong, Dong and Boroyevich, Dushan and Burgos, Rolando and Mattavelli, Paolo and Shen, Zhiyu},
journal={IEEE Transactions on Power Electronics}, 
title={{Impedance-Based Analysis of Grid-Synchronization Stability for Three-Phase Paralleled Converters}}, 
year={2016},
volume={31},
number={1},
pages={26-38}
}

@ARTICLE{Impedance-Based_Analysis_DFIG,
author={Hu, Bin and Nian, Heng and Li, Meng and Xu, Yunyang and Liao, Yuming and Yang, Jun},
journal={IEEE Transactions on Industrial Electronics}, 
title={{Impedance-Based Analysis and Stability Improvement of DFIG System Within PLL Bandwidth}}, 
year={2022},
volume={69},
pages={5803-5814}
}

@INPROCEEDINGS{PLL_power_converter,
author={Freijedo, Francisco D. and Doval-Gandoy, Jesus and Lopez, Oscar and Martinez-Peñalver, Carlos and Yepes, Alejandro G. and Fernandez-Comesaña, Pablo and Malvar, Jano and Nogueiras, Andres and Marcos, Jorge and Lago, Alfonso},
booktitle={2009 35th Annual Conference of IEEE Industrial Electronics}, 
title={{Grid-Synchronization Methods for Power Converters}}, 
year={2009},
volume={},
number={},
pages={522--529}
}

@ARTICLE{Small-Signal_Stability_Analysis_Based_Measured,
author={Wen, Bo and Boroyevich, Dushan and Burgos, Rolando and Mattavelli, Paolo and Shen, Zhiyu},
journal={IEEE Transactions on Power Electronics}, 
title={{Small-Signal Stability Analysis of Three-Phase AC Systems in the Presence of Constant Power Loads Based on Measured d-q Frame Impedances}}, 
year={2015},
volume={30},
number={10},
pages={5952-5963}
}

@article{IET_NFP,
author = {Harrison, Sam and Henderson, Callum and Papadopoulos, Panagiotis N. and Egea-Alvarez, Agusti},
title = {{Demystifying Inertial Specifications; Supporting the Inclusion of Grid-Followers}},
journal = {IET Renewable Power Generation},
volume = {17},
number = {7},
pages = {1768-1782},
year = {2023}
}

@ARTICLE{XiongfeiReferenceFrame,
author={Wu, Yang and Wu, Heng and Zhao, Fangzhou and Zhou, Zichao and Wang, Xiongfei},
journal={IEEE Transactions on Power Electronics}, 
title={{Reference-Frame Selection on Impedance Modeling of VSCs With Fundamental Frequency Dynamics}}, 
year={2025},
volume={40},
number={7},
pages={10059--10076}
}

@misc{GC0137,
author = {{National Grid ESO}},
title = {{GC013: Minimum Specification Required for Provision of GB Grid Forming Capability (Formerly Virtual Synchronous Machine/VSM Capability)}},
publisher = {National Grid ESO},
year = {2021}
}

@article{DFIG_PLL_IJEPES,
title = {{An Improved Virtual Inductance Control Method Considering PLL Dynamic Based on Impedance Modeling of DFIG under Weak Grid}},
journal = {International Journal of Electrical Power \& Energy Systems},
volume = {118},
pages = {105772},
year = {2020},
author = {Xueguang Zhang and Yage Zhang and Ran Fang and Dianguo Xu}
}

@article{chakraborty2022review,
title={{A Review of Active Probing-Based System Identification Techniques with Applications in Power Systems}},
author={Chakraborty, Rahul and Jain, Himanshu and Seo, Gab-Su},
journal={International Journal of Electrical Power \& Energy Systems},
volume={140},
pages={108008},
year={2022},
publisher={Elsevier}
}

@ARTICLE{DFIG_Frequency_Disturbances,
author={Kayikci, Mustafa and Milanovic, Jovica V.},
journal={IEEE Transactions on Power Systems}, 
title={{Dynamic Contribution of DFIG-Based Wind Plants to System Frequency Disturbances}}, 
year={2009},
volume={24},
number={2},
pages={859--867}
}

@article{Effect_in_FreqDynamics,
author = {Adrees, Atia and Milanović, J.V. and Mancarella, Pierluigi},
title = {{Effect of Inertia Heterogeneity on Frequency Dynamics of Low-Inertia Power Systems}},
journal = {IET Generation, Transmission \& Distribution},
volume = {13},
number = {14},
pages = {2951--2958},
year = {2019}
}

@book{ogata2010modern,
  title={Modern Control Engineering (fifth edition)},
  author={Ogata, Katsuhiko},
  year={2010}
}

@book{milano2020eigenvalue,
  title={Eigenvalue Problems in Power Systems},
  author={Milano, Federico and Dassios, Ioannis and Liu, Muyang and Tzounas, Georgios},
  year={2020},
  publisher={CRC Press}
}

@article{amin2017small,
title={{Small-Signal Stability Assessment of Power Electronics Based Power Systems: A Discussion of Impedance-and Eigenvalue-Based Methods}},
author={Amin, Mohammad and Molinas, Marta},
journal={IEEE Transactions on Industry Applications},
volume={53},
number={5},
pages={5014--5030},
year={2017}
}

@article{mudalige2024control,
title={{On Control Loop Interaction and EMT Dynamics in Shaping the Network Frequency Perturbation Plot}},
author={Mudalige, Anuradha and Wu, Heng and Langwasser, Marius and Liserre, Marco},
journal={IEEE Journal of Emerging and Selected Topics in Industrial Electronics},
year={2024},
publisher={IEEE}
}

@article{wei2025unified,
title={{A Unified Analysis Method for Harmonic and Frequency/Voltage Stability of Inverter-Integrated Power Systems}},
author={Wei, Fengting and Zhang, Haitao and Wang, Xiuli and Saeedifard, Maryam and Wang, Ziang and Wang, Xifan},
journal={IEEE Transactions on Power Systems},
year={2025}
}

@article{cecati2025interoperability,
title={{Interoperability Specifications for Multi-Vendor Converter-Dominated Grid: A Robust Stability Perspective}},
author={Cecati, Federico and Liserre, Marco},
journal={IEEE Transactions on Smart Grid},
year={2025},
volume={16},
number={4},
pages={3003--3016},
}

@inproceedings{milano2018foundations,
title={{Foundations and Challenges of Low-Inertia Systems}},
author={Milano, Federico and D{\"o}rfler, Florian and Hug, Gabriela and Hill, David J and Verbi{\v{c}}, Gregor},
booktitle={2018 power systems computation conference (PSCC)},
pages={1--25},
year={2018}
}

@article{moutevelis2024virtual,
title={{Virtual Synchronous Machine Design for Islanded Microgrids Using the Extended Impedance Criterion with Grid Frequency Dynamics Included}},
author={Moutevelis, Dionysios and Rold{\'a}n-P{\'e}rez, Javier and Rodr{\'\i}guez-Ortega, Pablo and Prodanovi{\'c}, Milan},
journal={IEEE Transactions on Energy Conversion},
year={2024}
}

@article{li2022systematic,
title={{A Systematic Stability Enhancement Method for Microgrids with Unknown-Parameter Inverters}},
author={Li, Yang and Wu, Xiangyang and Shuai, Zhikang and Zhou, Quan and Chen, Haojie and Shen, Zheng John},
journal={IEEE Transactions on Power Electronics},
volume={38},
number={3},
pages={3029--3043},
year={2022}
}

@article{wu2024influence,
title={{Influence of PLL on Stability of Interconnected Grid-Forming and Grid-Following Converters}},
author={Wu, Yang and Wu, Heng and Zhao, Fangzhou and Li, Zejie and Wang, Xiongfei},
journal={IEEE Transactions on Power Electronics},
volume={39},
number={10},
pages={11980--11985},
year={2024}
}

@ARTICLE{11224594,
author={Alican, Onur and Moutevelis, Dionysios and Arévalo-Soler, Josep and Collados-Rodriguez, Carlos and Amorós, Jaume and Gomis-Bellmunt, Oriol and Cheah-Mañe, Marc and Prieto-Araujo, Eduardo},
journal={IEEE Transactions on Power Systems}, 
title={{A Dynamic Similarity Index for Assessing Voltage Source Behaviour in Power Systems}}, 
year={2025},
volume={},
number={},
pages={1-14}
}

@article{arevalo2025matlab,
title={{A Matlab-Based Toolbox for Automatic EMT Modeling and Small-Signal Stability Analysis of Modern Power Systems}},
author={Arevalo-Soler, Josep and Moutevelis, Dionysios and Mateu-Barriendos, Elia and Alican, Onur and Collados-Rodriguez, Carlos and Cheah-Ma{\~n}e, Marc and Prieto-Araujo, Eduardo and Gomis-Bellmunt, Oriol},
journal={arXiv preprint arXiv:2506.22201},
year={2025}
}

@misc{STAMP_Github,
author = {CITCEA-UPC},
title= {{{STAMP}: A Small-Signal Toolbox for Analysis of Modern Power systems}},
year = {2025},
howpublished = {\url{https://github.com/CITCEA-UPC/STAMP_Public.git}},
}

@article{tomas2025vector,
title={{A Vector-Based Flexible-Complexity Tool for Simulation and Small-Signal Analysis of Hybrid AC/DC Power Systems}},
author={Tom{\'a}s-Mart{\'\i}n, Andr{\'e}s and Zuluaga-R{\'\i}os, Carlos David and Su{\'a}rez-Porras, Jorge and Garc{\'\i}a-Aguilar, Javier and Sigrist, Lukas and Garc{\'\i}a-Cerrada, Aurelio and Kazemtabrizi, Behzad},
journal={Sustainable Energy, Grids and Networks},
pages={101817},
year={2025},
 publisher={Elsevier}
}

@article{bouterakos2025eigenvalue,
title={{On the Eigenvalue Tracking of Large-Scale Systems}},
author={Bouterakos, Andreas and McKeon, Joseph and Tzounas, Georgios},
journal={International Journal of Electrical Power \& Energy Systems},
volume={172},
pages={111310},
year={2025},
publisher={Elsevier}
}

@article{sun2011impedance,
title={{Impedance-Based Stability Criterion for Grid-Connected Inverters}},
author={Sun, Jian},
journal={IEEE transactions on power electronics},
volume={26},
number={11},
pages={3075--3078},
year={2011},
publisher={IEEE}
}

@article{sun2022frequency,
title={{Frequency-Domain Stability Criteria for Converter-Based Power Systems}},
author={Sun, Jian},
journal={IEEE Open Journal of Power Electronics},
volume={3},
pages={222--254},
year={2022},
publisher={IEEE}
}

@article{garcia2026siad,
title={{SIaD-Tool: A Comprehensive Frequency-Domain Tool for Small-Signal Stability and Interaction Assessment in Modern Power Systems}},
author={Garcia-Reyes, Luis A and Gomis-Bellmunt, Oriol and Prieto-Araujo, Eduardo and Lacerda, Vin{\'\i}cius A and Cheah-Ma{\~n}e, Marc},
journal={arXiv preprint arXiv:2601.05519},
year={2026}
}

@article{garcia2026z,
title={{Z-Tool: Frequency-Domain Characterization of EMT Models for Small-Signal Stability Analysis}},
author={Garcia, Francisco Javier Cifuentes and Beerten, Jef},
journal={Electric Power Systems Research},
volume={252},
pages={112405},
year={2026},
publisher={Elsevier}
}

@article{zhou2024rapid,
title={{Rapid Admittance Measurement of Power Converters using Double-PLL Grid-Following Inverters}},
author={Zhou, Weihua and Ravanji, Mohammad Hasan and Mohammed, Nabil and Bahrani, Behrooz},
journal={IEEE Transactions on Power Delivery},
volume={39},
number={3},
pages={1407--1419},
year={2024},
publisher={IEEE}
}

@inproceedings{haberle2023mimo,
title={{MIMO Grid Impedance Identification of Three-Phase Power Aystems: Parametric vs. Nonparametric Approaches}},
author={Haberle, Verena and Huang, Linbin and He, Xiuqiang and Prieto-Araujo, Eduardo and Smith, Roy S and Dorfler, Florian},
booktitle={2023 62nd IEEE Conference on Decision and Control (CDC)},
pages={542--548},
year={2023},
organization={IEEE}
}

@ARTICLE{11328961,
author={Cheah-Mane, Marc and Arevalo-Soler, Josep and Moutevelis, Dionysios and Mateu-Barriendos, Elia and Prieto-Araujo, Eduardo and Gomis-Bellmunt, Oriol and Renedo, Javier and Martin-Almenta, Macarena and Nuño-Martinez, Edgar and Martinez-Villanueva, Sergio and Jung, JooYong and Kim, Namkyu and Kwon, YoungJin},
journal={IEEE Power and Energy Magazine}, 
title={{A New Paradigm for Small-Signal Stability Analysis in Modern Power Systems: Challenges, Models, and Methods in Power Systems Rich in Power Electronics}}, 
year={2026},
volume={24},
number={1},
pages={67--81}
}

@article{benedetti2025exploring,
title={{Exploring the Variation of Critical Modal Properties in Power Systems With Converter-Interfaced Units}},
author={Benedetti, Luke Ian and Papadopoulos, Panagiotis N and Preece, Robin},
journal={Sustainable Energy, Grids and Networks},
pages={101841},
year={2025},
publisher={Elsevier}
}

@article{desai2026effect,
title={{Effect of Dispatch Decisions on Small-Signal Stability of Converter-Dominated Power Systems}},
author={Desai, Maitraya Avadhut and Stanojev, Ognjen and Muntwiler, Simon and Hug, Gabriela},
journal={arXiv preprint arXiv:2601.05070},
year={2026}
}

@inproceedings{smith2024black,
title={{Black-Boxing of Converter State-Space Models for Power System Eigenvalue Analysis}},
author={Smith, Andrew Macmillan and Tutturen, Trond and D’Arco, Salvatore and Suul, Jon Are},
booktitle={2024 IEEE PES Innovative Smart Grid Technologies Europe (ISGT EUROPE)},
pages={1--5},
year={2024}
}

@inproceedings{Bollerslev2025RootCause,
author = {Bollerslev, J. and Liao, Y. and Wu, H. and Kwon, J. and Wang, X.},
title = {{Root Cause Identification of Small-Signal Instabilities of Converter-Based Power Systems}},
booktitle = {Proceedings of the 24th Wind and Solar Integration Workshop},
year = {2025},
}

@inproceedings{arbogast2025impact,
title={{Impact from Black-Box Small-Signal State-Space Representation on Participation Factors Analysis}},
author={Arbogast, Louis and Tutturen, Trond and D’Arco, Salvatore and Suul, Jon Are and Paolone, Mario},
booktitle={IECON 2025--51st Annual Conference of the IEEE Industrial Electronics Society},
pages={1--8},
year={2025}
}

@article{mohammed2022comparison,
title={{Comparison of PLL-based and PLL-less Control Strategies for Grid-Following Inverters Considering Time and Frequency Domain Analysis}},
author={Mohammed, Nabil and Zhou, Weihua and Bahrani, Behrooz},
journal={IEEE Access},
volume={10},
pages={80518--80538},
year={2022},
publisher={IEEE}
}

@article{mohammed2022fast,
title={{Fast and Accurate Grid Impedance Estimation Approach for Stability Analysis of Grid-Connected Inverters}},
author={Mohammed, Nabil and Ciobotaru, Mihai},
journal={Electric Power Systems Research},
volume={207},
pages={107831},
year={2022},
publisher={Elsevier}
}

@article{mohammed2019online,
title={{Online Parametric Estimation of Grid Impedance under Unbalanced Grid Conditions}},
author={Mohammed, Nabil and Ciobotaru, Mihai and Town, Graham},
journal={Energies},
volume={12},
number={24},
pages={4752},
year={2019},
publisher={MDPI}
}

@article{mohammed2022online,
title={{Online Grid Impedance Estimation-Based Adaptive Control of Virtual Synchronous Generators Considering Strong and Weak Grid Conditions}},
author={Mohammed, Nabil and Ravanji, Mohammad Hasan and Zhou, Weihua and Bahrani, Behrooz},
journal={IEEE Transactions on Sustainable Energy},
volume={14},
number={1},
pages={673--687},
year={2022},
publisher={IEEE}
}

@article{NabilFrequencyEstimation,
author = {Mohammed, Nabil and Ciobotaru, Mihai and Town, Graham},
title = {{Fundamental Grid Impedance Estimation Using Grid-Connected Inverters: A Comparison of two Frequency-Based Estimation Techniques}},
journal = {IET Power Electronics},
volume = {13},
number = {13},
pages = {2730--2741},
year = {2020}
}

@article{reissner2024robust,
title={{Robust and Adaptive Tuning of PI Current Controllers for Grid Forming Inverters}},
author={Rei{\ss}ner, Florian and Weiss, George},
journal={IEEE Open Journal of the Industrial Electronics Society},
year={2024},
publisher={IEEE}
}

@article{zhu2021participation,
title={{Participation Analysis in Impedance Models: The Grey-Box Approach for Power System Stability}},
author={Zhu, Yue and Gu, Yunjie and Li, Yitong and Green, Timothy C},
journal={IEEE Transactions on Power Systems},
volume={37},
number={1},
pages={343--353},
year={2021},
publisher={IEEE}
}

@article{yang2021siso,
title={{SISO Impedance-Based Stability Analysis for System-Level Small-Signal Stability Assessment of Large-Scale Power Electronics-Dominated Power Systems}},
author={Yang, Dongsheng and Sun, Yin},
journal={IEEE Transactions on Sustainable Energy},
volume={13},
number={1},
pages={537--550},
year={2021},
publisher={IEEE}
}

@article{chen2023small,
title={{Small-Signal Stability Criteria in Power Electronics-Dominated Power Systems: a Comparative Review}},
author={Chen, Qifan and Bu, Siqi and Chung, Chi Yung},
journal={Journal of Modern Power Systems and Clean Energy},
volume={12},
number={4},
pages={1003--1018},
year={2023},
publisher={SGEPRI}
}

@article{zhang2019impedance,
title={{Impedance-Based Analysis of Interconnected Power Electronics Systems: Impedance Network Modeling and Comparative Studies of Stability Criteria}},
author={Zhang, Chen and Molinas, Marta and Rygg, Atle and Cai, Xu},
journal={IEEE Journal of Emerging and Selected Topics in Power Electronics},
volume={8},
number={3},
pages={2520--2533},
year={2019},
publisher={IEEE}
}

@article{rygg2017apparent,
title={{Apparent Impedance Analysis: A Small-Signal Method for Stability Analysis of Power Electronic-Based Systems}},
author={Rygg, Atle and Molinas, Marta},
journal={IEEE Journal of Emerging and Selected Topics in Power Electronics},
volume={5},
number={4},
pages={1474--1486},
year={2017},
publisher={IEEE}
}

@INPROCEEDINGS{11305437,
author={Alican, Onur and Moutevelis, Dionysios and Arévalo-Soler, Josep and Gomis-Bellmunt, Oriol and Cheah-Mañe, Marc and Prieto-Araujo, Eduardo},
booktitle={2025 IEEE PES Innovative Smart Grid Technologies Conference Europe (ISGT Europe)}, 
title={{A Small-Signal Framework for Active Power-to-Frequency Dynamics Analysis in Transmission Networks}}, 
year={2025},
volume={},
number={},
pages={1-5}
}

@article{milano2021complex,
title={Complex Frequency},
author={Milano, Federico},
journal={IEEE Transactions on Power Systems},
volume={37},
number={2},
pages={1230--1240},
year={2021}
}

@article{moutevelis2023taxonomy,
title={{Taxonomy of Power Converter Control Schemes Based on the Complex Frequency Concept}},
author={Moutevelis, Dionysios and Rold{\'a}n-P{\'e}rez, Javier and Prodanovic, Milan and Milano, Federico},
journal={IEEE Transactions on Power Systems},
volume={39},
number={1},
pages={1996--2009},
year={2023},
publisher={IEEE}
}
\end{spacing}
\endgroup

\bibliographystyle{IEEEtran}
\end{document}